\documentclass{article}
\pdfoutput=1 
\usepackage{microtype}
\usepackage{graphicx}
\usepackage{subcaption}
\usepackage{booktabs} 
\usepackage{amsfonts}
\usepackage{amsmath}
\usepackage{enumitem}
\usepackage{soul}
\usepackage{etoolbox}

\usepackage{stackrel,amssymb}
\newcommand{\leftrarrows}{\mathrel{\raise.75ex\hbox{\oalign{%
  $\scriptstyle\leftarrow$\cr
  \vrule width0pt height.5ex$\hfil\scriptstyle\relbar$\cr}}}}
\newcommand{\lrightarrows}{\mathrel{\raise.75ex\hbox{\oalign{%
  $\scriptstyle\relbar$\hfil\cr
  $\scriptstyle\vrule width0pt height.5ex\smash\rightarrow$\cr}}}}
\newcommand{\Rrelbar}{\mathrel{\raise.75ex\hbox{\oalign{%
  $\scriptstyle\relbar$\cr
  \vrule width0pt height.5ex$\scriptstyle\relbar$}}}}
\newcommand{\longleftrightarrows}{\leftrarrows\joinrel\Rrelbar\joinrel\lrightarrows}

\makeatletter
\def\leftrightarrowsfill@{\arrowfill@\leftrarrows\Rrelbar\lrightarrows}
\newcommand{\xleftrightarrows}[2][]{\ext@arrow 3399\leftrightarrowsfill@{#1}{#2}}
\makeatother


\usepackage{relsize}

\usepackage{lineno}

\newcommand\xo{0.0}
\newcommand\yo{0.0}
\newcommand\dx{1.25}
\newcommand\dy{1}

\newsavebox{\tempbox}

\newcommand{\textbox}[1]
{\savebox{\tempbox}{#1}
 \ifdim\wd\tempbox<2cm\relax
   \makebox[2cm]{\usebox{\tempbox}}%
 \else
   \parbox{2cm}{\centering #1}%
 \fi}

\usepackage{amsthm}

\newtheorem*{remark}{Hypothesis}

\newcommand{\norm}[1]{\left\lVert#1\right\rVert}

\usepackage[dvipsnames]{xcolor}

\newcommand{\pathToPlotImages}{./images-low}

\usepackage {tikz}
\usepackage[font=scriptsize,skip=0pt]{subcaption}
\usetikzlibrary {positioning}
\usetikzlibrary{shapes,snakes}
\usetikzlibrary{arrows,arrows.meta}
\usepackage{tikz-cd}
\pgfdeclarelayer{background}
\pgfdeclarelayer{foreground}
\pgfsetlayers{background,main,foreground}
\tikzset{fontscale/.style = {font=\relsize{#1}}}
\usetikzlibrary{patterns}
\usepackage{hyperref}
\usepackage{varwidth}

\usepackage[skip=0pt]{caption}

\usepackage{setspace}

\newcommand{\Sp}{\upsilon}
\newcommand{\Sq}{\rho}

\newcommand*{\boxcolorBlue}{NavyBlue!10}
\makeatletter
\newcommand{\boxedBlue}[1]{\textcolor{\boxcolorBlue}{%
\tikz[baseline={([yshift=-1ex]current bounding box.center)}] \node [rectangle,inner sep=0pt, outer sep=0pt, fill=NavyBlue!10,rounded corners,draw] {\normalcolor\m@th$\displaystyle#1$};}}
\makeatother

\newcommand*{\boxcolorGreen}{ForestGreen!10}
\makeatletter
\newcommand{\boxedGreen}[1]{\textcolor{\boxcolorGreen}{%
\tikz[baseline={([yshift=-1ex]current bounding box.center)}] \node [rectangle,inner sep=0pt, outer sep=0pt, fill=ForestGreen!10,rounded corners,draw] {\normalcolor\m@th$\displaystyle#1$};}}
\makeatother

\newcommand*{\boxcolorRed}{Red!30}
\makeatletter
\newcommand{\boxedRed}[1]{\textcolor{\boxcolorRed}{%
\tikz[baseline={([yshift=-1ex]current bounding box.center)}] \node [circle, inner sep=0pt, outer sep=0pt,fill=Red!30,rounded corners,draw] {\normalcolor\m@th$\displaystyle#1$};}}
\makeatother

\usepackage[accepted]{icml2021}

\icmltitlerunning{GEM: Group Enhanced Model for Learning Dynamical Control Systems}

\begin{document}

\twocolumn[
\icmltitle{GEM: Group Enhanced Model for Learning Dynamical Control Systems}



\icmlsetsymbol{equal}{*}

\begin{icmlauthorlist}
\icmlauthor{Philippe Hansen-Estruch}{UCB}
\icmlauthor{Wenling Shang}{UA}
\icmlauthor{Lerrel Pinto}{NYU}
\icmlauthor{Pieter Abbeel}{UCB}
\icmlauthor{Stas Tiomkin}{UCB}
\end{icmlauthorlist}

\icmlaffiliation{UCB}{BAIR Lab, University of California, Berkeley}
\icmlaffiliation{UA}{University of Amsterdam}
\icmlaffiliation{NYU}{New York University}

\icmlcorrespondingauthor{Stas Tiomkin}{stas@berkeley.edu}

\icmlkeywords{Machine Learning, ICML}

\vskip 0.3in
]



\printAffiliationsAndNotice{} 


{

\begin{abstract}
Learning the dynamics of a physical system wherein an autonomous agent operates is an important task. Often these systems present apparent geometric structures.
For instance, the trajectories of a robotic manipulator can be broken down into a collection of its transitional and rotational motions, fully characterized by the corresponding Lie groups and Lie algebras. In this work, we take advantage of these structures to build effective dynamical models that are amenable to sample-based learning. We hypothesize that learning the dynamics on a Lie algebra vector space is more effective than learning a direct state transition model. To verify this hypothesis, we introduce the \emph{Group Enhanced Model} (GEM). GEMs significantly outperform conventional transition models on tasks of long-term prediction, planning, and model-based reinforcement learning across a diverse suite of standard continuous-control environments, including Walker, Hopper, Reacher, Half-Cheetah, Inverted Pendulums, Ant, and Humanoid. { Furthermore, plugging GEM into existing state of the art systems enhances their performance, which we demonstrate on the PETS system.} This work sheds light on a connection between learning of dynamics and Lie group properties, which opens doors for new research directions and practical applications along this direction. Our code is publicly available at: https://tinyurl.com/GEMMBRL.
\end{abstract}

\section{Introduction}\label{sec:Intro}
The interaction between an autonomous agent and its environment can be modeled by a dynamical control system, where the environment's current state and agent's action are mapped by a dynamics function to the next state~\citep{watter2015embed,todorov2005generalized}. When the system is unknown and complex, learning a reliable dynamical model from data samples is a notably challenging problem. 

Reliable dynamical models are required in many important applications such as planning~\citep{schrittwieser2020mastering}, model-based reinforcing learning~\citep{moerland2020model}, system identification~\citep{nelles2020nonlinear}, environment emulation~\citep{castelletti2012data}, SimToReal ~\cite{tan2018sim}, etc. Moreover, reliably predicting over long-time horizons remains a brutishly difficult problem as compared to a single-step prediction. Such long horizon prediction is particularly useful for downstream planning of robust behaviors \cite{chua2018deep,rajeswaran2020game,hafner2020mastering}. Hence, there has been a growing research interest in enhancing the quality of sample-based methods for learning dynamical models~\citep{brzeski2017sample, fragkiadaki2015recurrent, deisenroth2011learning}.    



\begin{figure}[t!]
\centering
\includegraphics[scale=0.45]{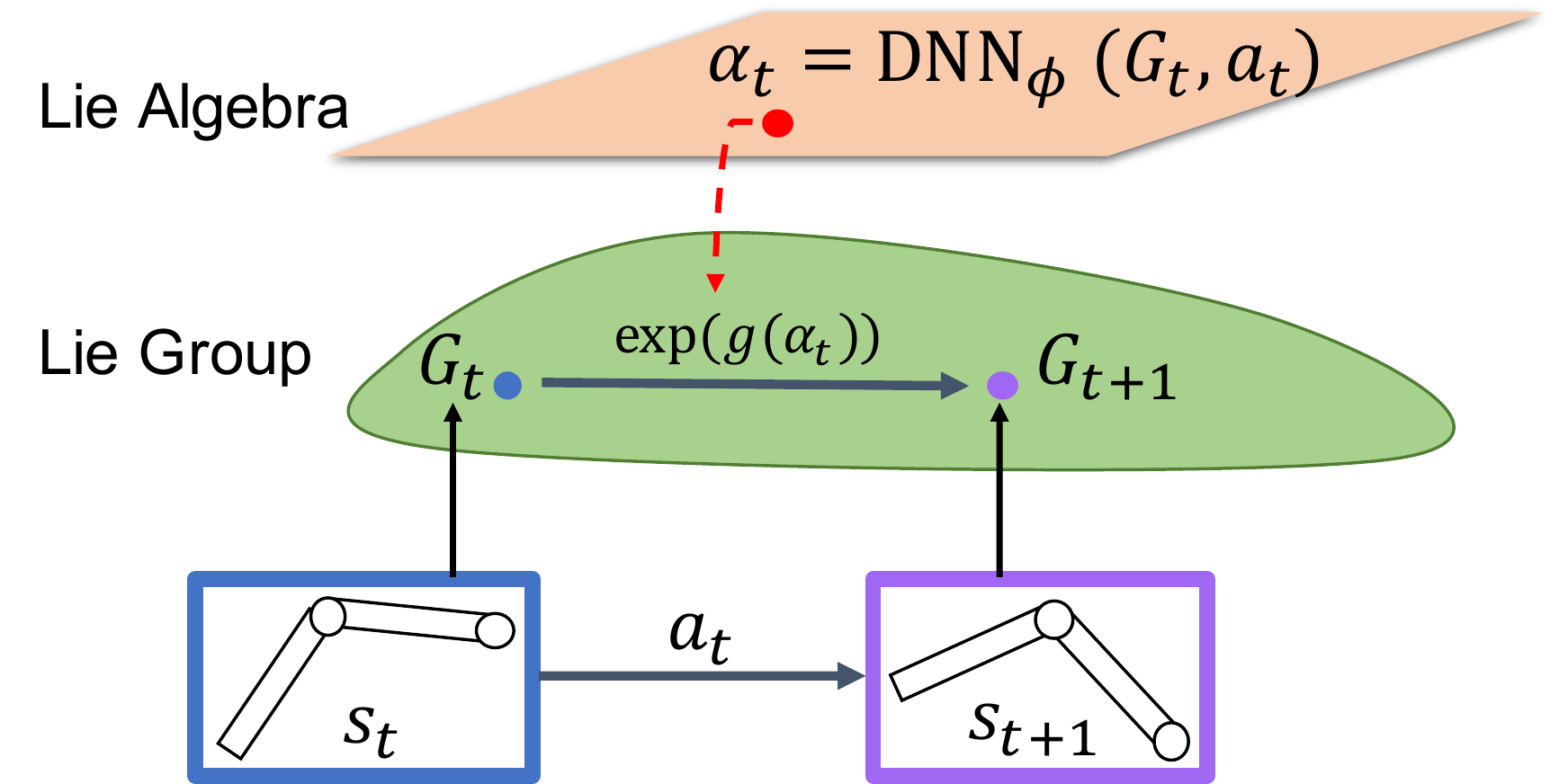}
\caption{{An overview of our proposed GEM. We convert the raw state $s_t$ to a Lie group representation, $G_t$. GEM uses a deep neural net to predict the Lie algebra, $\alpha_t$, that evolves $G_t$ via the exponential map to the next group element, $G_{t+1}$, corresponding to the raw state $s_{t+1}$, obtained by applying action $a_t$ to $s_t$.}}\label{fig:teaser}
\end{figure}

In general, the problem of learning a dynamical model is formulated as learning a predictive mapping from the present state and action to the future state~\citep{schrittwieser2020mastering}. Usually, the outputs and the inputs in learning of this mapping are treated as general Euclidean vectors. However, such a view overlooks the intrinsic properties within the state space of a dynamical system. In fact, the states form a manifold structure~\citep{strogatz2018nonlinear,agrachev2013control}. In this work we propose a novel approach to exploit this underlying structure and learn reliable dynamical models. 

An important component for sample-based learning of dynamical models is the choice of priors, which usually comes from our understanding of the underlying physical laws~\citep{NEURIPS2019_26cd8eca, lnn2020}. For example, the priors can be a prior distribution over the model parameters~\citep{chua2018deep, luo2020bayesian} or a set of differential equations with undetermined physical parameters~\citep{bahl2020neural, chen2018neural}. Here, we consider another type of priors, which are \textit{primitive geometric motions} of dynamical system. These priors reflect the manifold structure of the state space of dynamical system, as explained below.  

Primitive geometric priors describe \textit{what a dynamical system does}, which is independent from any specific physical parameters (such as friction, mass, natural frequencies, etc) or physical laws. For example, by simply observing interactions between the agents and the environments, one may understand that a car rotates and translates linearly, a robotic arm rotates around rotation axes of its joints, and a humanoid translates its center of mass while rotating its hinge joints. Such information can  be obtained either implicitly via cameras \cite{SymmetryNet,gothandaraman2020reflectional} or explicitly via hardware specifications~\citep{todorov2012mujoco} with little extra cost. In contrast, many physical states  need to be measured by elaborated motion capture or sensory systems~\citep{OpenAIHand}.

To characterize these geometric priors, we follow the framework of Lie group theory, a powerful tool for studying manifolds with group structure \cite{hall2003lie}\footnote{This theory is well-known dating back to the XIX century, when mathematician Sophus Lie laid the foundations of the theory of continuous transformation groups.}. To this end, we propose the \emph{Group Enhanced Models} (GEMs) that incorporates the properties of the smooth manifold, which allows a neural network to effectively learn reliable dynamical models both in short and long horizons. 

Central to GEMs is the one-to-one correspondence between elements from the manifold and those from an associated linear vector space, termed Lie Algebra. The significance of this correspondence is summarized in the following paragraph by \citet{howe1983very}:
\vspace{-0.25cm}
\begin{quotation}
{\it
\noindent ``...The essential phenomenon of Lie theory is that one may associate in a natural way to a Lie group $\mathcal{G}$ its Lie algebra $\mathfrak{g}$. Thus for many purposes one can replace $\mathcal{G}$ with g. Since $\mathcal{G}$ is a complicated nonlinear object and $\mathfrak{g}$ is just a vector space, it is usually vastly simpler to work with $\mathfrak{g}$..."} 
\end{quotation}
\begin{remark}\label{eq:Hypothesis}
We hypothesize that it is more efficient and effective to learn a dynamical system on the linear vector space Lie algebra, $\mathfrak{g}$, than on the original unstructured state space, $\mathcal{S}$, or on the complicated Lie group, $\mathcal{G}$.
\end{remark}

To explore this hypothesis, we develop GEM to enable learning a dynamical system on the linear vector space of Lie algebra. Figure~\ref{fig:teaser} provides an overview of GEM: after converting a raw state to a Lie group manifold, GEM outputs a Lie algebra action to evolve it to the target next state on the manifold that is equivalent to applying an action in the original raw space. In experiments, we evaluate GEM on various dynamical systems in different standard benchmark tasks. Our main contributions are:
\begin{itemize}[noitemsep,topsep=0pt]
    \item We propose a new framework, the \emph{Group Enhanced Models}, for learning dynamical models on Lie algebra.
    \item We show that GEM outperforms the relevant  baselines on the task of \textit{long horizon prediction} for standard continuous control benchmarks.
    \item We demonstrate the advantage of GEM for offline planning and model-based reinforcement learning on various dynamical systems. 
\end{itemize}              
In Section \ref{sec:relevant}, we review the related prior works. Section~\ref{sec:Preliminaries} overviews the relevant background of Lie group theory. Section \ref{sec:Method} introduces our method. Section \ref{sec:mainresult} presents the experimental results. Finally, Section \ref{sec:discussion} summaries and provides a broader perspective.

\section{Related Works}\label{sec:relevant}
In this section, we review the related prior works, including generic methods of learning dynamics, dynamical models with physical priors and machine learning models with Lie theory elements. 

\subsection{Generic predictive models}
The most common technique for learning dynamics is to fit either the forward model, $s_{t+1}=f_{\theta}(s_t, a_t)$, or the inverse-model, $a_t = f^{-1}_{\theta}(s_{t+1}, s_t)$ to the training samples of $N$-step state-action trajectories, $\tau=\{s_t, a_t\}_{t=1}^N$ \cite{nguyen2010using,nguyen2011model,schaal2002scalable,kocijan2004gaussian,deisenroth2011learning}. 

Feed-forward networks are commonly used model classes to parametrize these forward models \cite{NEURIPS2019_5faf461e}.
%
The training objective is generally formulated as empirical loss minimization between the predicted next state and the true next state, given the present state and agent's action:
\begin{align}
    \underset{\theta}{\arg\min}\frac{1}{N}\sum_{t=1}^N\; loss(f_{\theta}(s_t, a_t), s_t), 
\end{align}
\noindent where $f_{\theta}(s_t, a_t)$, and $loss(\cdot, \cdot)$ are the predicted next state, $\hat{s}_{t+1}$, and an appropriate loss function, such as mean squared error, respectively. 

This generic approach manages to achieve decent results, and it is often adopted as the comparative baseline when proposing other novel approaches~\citep{NEURIPS2019_26cd8eca,lnn2020,sanchez2018graph,lutter2019deep}. 
Here, we also compare our proposed methods to the generic baseline.

\subsection{Dynamical models with physical priors}\label{sec:Dynamical models with physical priors}
Recently, a series of works have proposed to incorporate physical priors to model dynamics. One line of research takes advantage of the properties of the Hamiltonian and the Lagrangian function of dynamical systems \citep{NEURIPS2019_26cd8eca,lnn2020,lutter2019deep}. Specifically, \citet{NEURIPS2019_26cd8eca} propose  \textit{Hamiltonian Neural Networks} (HNN) to explicitly use the Hamiltonian formalism and the canonical coordinates.
The resulting \textit{Hamiltonian} optimization objective better preserves energy and hence improves long horizon predictions.

A follow-up idea, \textit{Lagrangian~Neural~Networks}~\citep{lnn2020} eliminates the need for the canonical coordinates.  An extended Lagrangian formalism, including external forces, is proposed by \citet{lutter2019deep} for learning inverse-models. These approaches improve the quality of sample-based learning of autonomous dynamical systems. 
In contrast, we consider dynamical control systems with external control signals. We also do not assume the knowledge of the Hamiltonian or Lagrangian formalisms but explore forward-dynamics models from the group theoretical viewpoint.   

Another approach to incorporate physical priors is proposed by~\citet{sanchez2018graph}, based on \textit{Graph Neural Networks} (GNN). The components of a dynamical system are embedded in a static and a dynamic graph with recurrent connections. The static graph represents the physical parameters (friction,  mass, viscosity etc), and the dynamic represents instantaneous state. Our method, on the other hand, does not need to explicitly learn these physical parameters.

\subsection{Models with Lie theory elements}
Several existing works have leveraged the power of Lie group theories in machine learning applications, from image classification~\citep{cohen2016group}, medical data preprocessing~\citep{bekkers2019b}, to action recognition~\cite{huang2017deep,vemulapalli2016rolling}, etc. 
The language of Lie theory is also ubiquitous in robotics~\citep{selig2004lie,byravan2017se3} and geometric optimal control~\citep{agrachev2013control}. 

A common theme of these prior works suggest that utilizing the Lie group representations instead of the original observations leads to better training efficiency and better final performance. 
In addition, \citet{quessard2020learning} learns disentangled representations by constraining the latent space to be a Lie group manifold. 
%
Similarly, \citet{byravan2017se3} employ the prediction of Lie group elements in the context of scene recognition and object segmentation for robotic manipulations, where the elements of $SE(3)$ are modeled by a neural network.


Unlike other machine learning frameworks using Lie theory elements, we consider both a Lie group manifold and its corresponding Lie algebra. In particular, the Lie algebra elements are modeled by a neural network and govern the transition between the present and the next points on the Lie group manifold. 



\section{Preliminaries}\label{sec:Preliminaries}

Our work considers rigid body dynamics~\citep{tsai1999robot}. Translations, rotations and their compositions are the primitive motions of such systems. This is a broad class of dynamical systems, used ubiquitously in artificial intelligence and optimal control \cite{todorov2012mujoco,brockman2016openai,selig2004lie}. For example, the motion of a planar single pendulum, which is a two-dimensional non-linear dynamical system, is characterized by rotation with regard to its pivot. For another example, the motion of a humanoid, which is a high-dimensional system, is characterized by a composition of multiple limb rotations and translation of center of mass. An important feature of these primitive motions is that they are elements of the corresponding groups, or more precisely, Lie groups. 

In this section, we first review the generic approach for sample-based dynamics learning and then review the relevant Lie group theories to our work.

\subsection{Generic predictive  dynamical models}\label{sec:generic}
{
A dynamics function, $f$, is a map from the current state, $s_i \in \mathcal{S}$ and the current action, $a_i\in \mathcal{A}$ to the following  state, $s_{i+1}\in \mathcal{S}$\footnote{in practice, following the previous works \cite{nagabandi2020deep}, we model the change in state, $\Delta \hat{s}_{i+1}=f(s_i, a_i)$.}:
\begin{align}\label{eq:dynamical model}
    s_{i+1} = f(s_i, a_i).
\end{align}
}
When the dynamics is unknown, learning an accurate $f$ becomes essential for e.g. planning and control~\citep{nagabandi2018neural, nagabandi2020deep, chua2018deep}. Usually, this map is approximated by a parametric function, $f_{\theta}$, through minimization of some loss between the predicted next state $\Delta \hat{s}_{i+1} = f_{\theta}(s_i, a_i)$ and the true next state $s_i$:
\begin{align}\label{eq:genericLoss}
    \theta^{*} = \underset{\theta}{\arg\min} \frac{1}{N}\sum_{i=0}^N loss(\hat{s}_{i+1}, s_{i+1}).
\end{align}
This generic way of learning dynamics often ignores any geometric structure within the dynamical system but simply operates on the unstructured raw state space $\mathcal{S}$. 


\subsection{Elements of Lie group theory}
A finite dimensional Lie group is a group and a differential manifold at the same time \cite{howe1983very,hall2003lie}.

For every point $x\in \mathcal{M}$ on the manifold, there exists a tangent linear vector space $\mathcal{TM}_{x}$. The tangent space at the identity element, $\mathcal{TM}_{\mathcal{E}}$, is special. It is called the Lie algebra, $\mathfrak{g}$, of the Lie group, $\mathcal{G}$. The Lie group (manifold) and its algebra (linear vector space) are connected by:
\begin{align}\label{eq:connection}
\mathfrak{g} \stackrel[\exp]{\log}\longleftrightarrows \mathcal{G},
\end{align}
where $\exp$ and $\log$ can be calculated via the corresponding Taylor series. For instance, $\forall g\in \mathfrak{g}$, $\exp(g) =\sum_{n=0}^{\infty}\frac{g^{(n)}}{n!}$, where $g^{(n)}{=}{g\circ g\circ\dots\circ  g}$ and $\circ$ is a binary composition operator defined on $\mathfrak{g}$.

The Lie algebra is a linear vector space spanned by a basis of $K$ elements, $E=\{E_1, E_2, \dots, E_K\}$, where $K$ is the dimension of the manifold $\mathcal{M}$. Every vector in the algebra, $g \in \mathfrak{g}$, can be represented by an unique linear combination of the basis elements with $K$ scalars, $\{\alpha_i\}_{i=1}^K$:
\begin{align}\label{eq:algebraelement}
    g(\alpha) = \sum_{i=1}^K \alpha_i E_i.
\end{align}
The equations \eqref{eq:connection} and \eqref{eq:algebraelement} read together as follows:
\begin{align}\label{eq:expTogroup}
    G(\alpha) = \exp(g(\alpha)) \in\mathcal{G}.
\end{align}
The composition axiom of group ensures that when a group element, e.g. $\exp(g_{\alpha})$, acts on another group element $G\in\mathcal{G}$, we obtain another element $G'$ within $\mathcal{G}$:
\begin{align}\label{eq:LieDynamics}
    G' = \exp(g(\alpha))\circ G.
\end{align}
In other words, the coefficients of Lie algebra govern the transition between $G$ and $G'$. This property is core to our proposed framework. Specifically, $G$ can be viewed as the Lie group representation of the current state and $G'$ the next state. In Section~\ref{sec:Method}, we will formally introduce our method for learning of dynamical control systems through Lie algebra coefficients, $\alpha$. 

\subsubsection{Groups of rotation and translation}
The Lie groups of 2D/3D spacial rotations, $SO(2)$/$SO(3)$, and translations, $SE(2)$/$SE(3)$, fully characterize the primitive geometric motion of rigid body, and are broadly used in robotics \cite{selig2004lie,bourmaud2015continuous,featherstone2014rigid}. Both groups admit matrix representations \cite{hall2003lie}. Here, the group composition operator, $\mathbf{\circ}$, is the standard matrix multiplication and $\exp$ the matrix exponential. We review the properties of these groups in the Supplementary Material. In this work, we assume the bases of $SO(3)$ and $SE(3)$ are known, which is a common setup in many robotic applications. In the case of unknown basis, it can potentially be learned as well~\cite{quessard2020learning}, which we leave for future investigation. {In contrast to~\cite{quessard2020learning}, we utilize known group structures as a prior for learning of dynamical model \eqref{eq:dynamical model} for planning.}

\textbf{Conversion from angle-axis representation} The elements of $SO(3)$ can be constructed from a rotation axis, $u$, and a rotation angle, $\theta$ \cite{selig2004lie}\footnote{{In the robotic environments we use in the current work, state is represented either by angle-axis or quaternion, which we firstly transform to the corresponding Lie groups, such as SO and or SE. The transformation from angle-axis, (which is used in most of the environments in this work), to SO is provided for completeness.}}:
\begin{align}\label{eq:convertion}
    G \!= \!\cos(\theta)\!\!\cdot\!\!\mathbf{I}\!+\! \sin(\theta)\!\!\cdot\!\!(u\times u) \!+\! (1\!-\!\cos(\theta))\!\!\cdot\!\!(u\otimes u),
\end{align}
where $\mathbf{I}$ is the identity matrix, $\times$ the cross product, and $\otimes$ the outer product. The axes of rotations can be revealed either from the hardware specification \cite{todorov2012mujoco}, or by computer vision techniques \cite{SymmetryNet,gothandaraman2020reflectional}.

\section{Our Method}\label{sec:Method}
Now, we are ready to introduce \emph{Group Enhanced Model} (GEM), which is a particular method for learning \eqref{eq:LieDynamics}. The core component in our method is prediction of Lie algebra coefficients from data samples that connect two consequent points on the manifold, $G_t$, and $G_{t+1}$. 
Our method includes a novel objective, given by equation \eqref{eq:LieLoss}, a two-staged neural network architecture, shown at Figure \eqref{fig:GEMscheme}, and its training algorithm, Algorithm \eqref{alg:GEM}. 

{\textbf{Notations:} Given a state observation, $s_t = (\Sq_t, \Sp_t)$, the first stage models the Lie algebra coefficients, $\alpha$, that  build the ``\textit{static}'' component, $\Sq_t$, and the second stage models the ``\textit{dynamic}'' component, $\Sp_t$, where ``\textit{static}'' and ``\textit{dynamic}'' represent position and velocity, or angle and angular velocity, respectively. The static components are represented by the corresponding group elements, {which can be any combination of $SO(2)$, $SE(2)$, $SO(3)$, $SE(3)$} \cite{hall2003lie}. {The full state vector is represented by stacking matrices of corresponding groups. For example, double pendulum is represented by stacking two rotation matrices. We refer to this collective group state as $G$.}}



\subsection{GEM Objective}\label{sec:AlgebraricLoss}
Our goal is to train a neural network to predict the Lie algebra coefficients, $\alpha_{i}$, to construct the next state by applying the associated group action $G(\alpha_{i})=\exp(g(\alpha_{i}))$ in \eqref{eq:expTogroup} to the current state. To do so, we formulate the following objective function where $\phi$ is the parameters of our network and $ {\hat{\alpha_{i}}} $ is its predicted coefficients :
\vspace{-0.2cm}
\begin{align}\label{eq:LieLoss}
    \phi^{*} =& \underset{\phi}{\arg\min} \frac{1}{N}\sum_{i=1}^N \tilde{l}\Bigl(\hat{G}_{i+1}, G_{i+1}\Bigr),\\
    =& \underset{\phi}{\arg\min} \frac{1}{N}\sum_{i=1}^N  \tilde{l}\Bigl(\exp\bigl(g(\hat{\alpha_{i}})\bigr)\circ G_i, G_{i+1}\Bigr),\nonumber
    \vspace{-0.2cm}
\end{align}
\noindent and we refer to $\tilde{l}(\cdot, \cdot)$ the ``\textit{algebraic loss}'', which depends on a particular representation of a group. In the case of a group with matrix representation, this loss is a matrix norm.


The loss in \eqref{eq:LieLoss} is used to train the model to predict the static components, $\Sq_t$, via the Lie algebra coefficients, $\alpha_t$. The velocity loss is \textit{mean square error} between true delta, $\Delta \Sp_t$, and predicted delta velocity, $\Delta \hat{\Sp}_t$, as shown in Algorithm \eqref{alg:GEM}. In Section \ref{sec:discussion} we discuss an alternative to this approach for predicting velocity.

\subsection{GEM Architecture}\label{sec:Arch}
In this section, we introduce an architecture for predicting the full state, including both static and dynamic state components. The architecture of GEM, shown in Figure \ref{fig:GEMscheme}, consists of two parts: Coefficient Model $\mathcal{L}^\alpha_\phi$ and Velocity Model $\mathcal{L}^{\Sp}_\psi$, and $\psi$. The Coefficient Model is a neural network, parameterized by $\phi$, with an arbitrary architecture that maps the Lie Group state $G_t$, the velocity $\Sp_t$, and the action $a_t$ at time $t$ into Lie Algebra coefficients $\alpha_t$ that describe the change of motion undergone by the primitive geometry. The Velocity Model, also an arbitrary neural network, parameterized by $\psi$, then maps $G_t$, $\Sp_t$, $\alpha_t$, and action $a_t$ to $\Delta \Sp_{t}$. From equation~\eqref{eq:algebraelement} and~\eqref{eq:LieDynamics}, we compute the final prediction of $G_{t+1}$ using $\alpha_t$ coefficients. We hypothesised that the predicted coefficients could aid in predicting future velocities, which we confirm by the experiments. { However, one can combine the coefficient model and velocity model into a single network without loss of the superiority of GEM. This is demonstrated by PETS comparison experiment in Section \ref{sec:gempets}.}

In the current work, we explore feed-forward networks for modeling of Lie algebra coefficients and velocities. However, other frameworks are possible, including recurrent networks, transformer networks, graph networks, etc, as we discuss in Section \ref{sec:discussion}. 


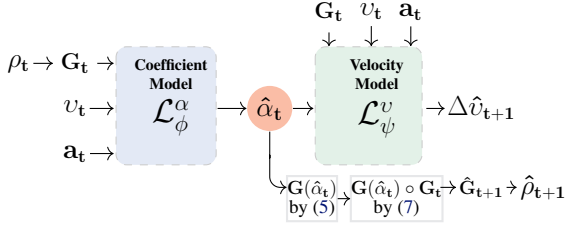
\begin{figure}[t!]
    \centering
    \newcommand{\myFillB}{White!25}
\begin{tikzpicture}[scale=.6, transform shape, every node/.style={thick,circle,inner sep=0pt},every text node part/.style={align=center}]

\tikzstyle{fontbf} = [draw,rectangle,text width=5cm,text centered,font=\bfseries]

\node[circle, draw=\myFillB,font=\fontsize{16}{0}\selectfont] (n01) at (\xo+0.3*\dx, \yo) {$\mathbf{\Sq_{t}}$};

\node[circle, draw=\myFillB,font=\fontsize{16}{0}\selectfont] (n02) at (\xo+1.3*\dx, \yo-\dy) {$\mathbf{\Sp_{t}}$};

\node[circle, draw=\myFillB,font=\fontsize{16}{0}\selectfont] (n03) at (\xo+1.3*\dx, \yo-2*\dy) {$\mathbf{a_{t}}$};

\node[regular polygon,regular polygon sides=9,draw=\myFillB,minimum size=1cm,font=\fontsize{15}{0}\selectfont] (n04) at (\xo+1.3*\dx, \yo) {$\mathbf{G_{t}}$};

\draw[->] (n01) -- (n04);

\node (n05) at (\xo+2.01*\dx, \yo) {};
\node (n06) at (\xo+2.01*\dx, \yo-\dy) {};
\node (n07) at (\xo+2.01*\dx, \yo-2*\dy) {};

\draw [->] (n02) -- (n06);
\draw [->] (n03) -- (n07);
\draw [->] (n04) -- (n05);

\node[circle, minimum size=1cm,font=\fontsize{20}{0}\selectfont] (n08) at (\xo+3*\dx, \yo-1.2*\dy) {${\mathbf{\mathcal{L}_{\phi}^{\alpha}}}$};

\node (n09) at (\xo+3.8*\dx, \yo-\dy) {};

\node[circle, fill=Red!30,minimum size=1cm,font=\fontsize{18}{0}\selectfont] (n10) at (\xo+4.75*\dx, \yo-\dy) {$\mathbf{\hat{\alpha}_{t}}$};

\draw [->] (n09) -- (n10);

\node (n11) at (\xo+5.55*\dx, \yo-\dy) {};
\node (n12) at (\xo+5.8*\dx, \yo+0.35*\dy) {};
\node (n13) at (\xo+6.55*\dx, \yo+0.35*\dy) {};
\node (n14) at (\xo+7.25*\dx, \yo+0.35*\dy) {};
\node (n15) at (\xo+7.5*\dx, \yo-\dy) {};

\draw [->] (n10) -- (n11);

\node[regular polygon,regular polygon sides=9,draw=\myFillB,font=\fontsize{15}{0}\selectfont] (n16) at (\xo+5.8*\dx, \yo+1.15*\dy) {$\mathbf{G_{t}}$};

\node[circle, draw=\myFillB,font=\fontsize{16}{0}\selectfont] (n17) at (\xo+6.55*\dx, \yo+1.15*\dy) {$\mathbf{\Sp_{t}}$};

\node[circle, draw=\myFillB,font=\fontsize{16}{0}\selectfont] (n18) at (\xo+7.25*\dx, \yo+1.15*\dy) {$\mathbf{a_{t}}$};

\draw [->] (n16) -- (n12);
\draw [->] (n17) -- (n13);
\draw [->] (n18) -- (n14);

\node[circle, minimum size=1cm,font=\fontsize{20}{0}\selectfont] (n19) at (\xo+6.65*\dx, \yo-1.25*\dy) {$\mathbf{\mathcal{L}_{\psi}^{\Sp}}$};


\node[circle, draw=\myFillB,font=\fontsize{16}{0}\selectfont] (n20) at (\xo+8.5*\dx, \yo-\dy) {$\Delta\mathbf{\hat{\Sp}_{t+1}}$};

\draw [->] (n15) -- (n20);

\node[rectangle, draw=Black!10, minimum size=1.0cm,font=\fontsize{12}{0}\selectfont ] (n21) at (\xo+5.5*\dx, \yo-3.0*\dy) {$\mathbf{G(\hat{\alpha}_t)}$\\by \eqref{eq:algebraelement}};
\node[rectangle, draw=Black!10, minimum size=1.0cm,font=\fontsize{12}{0}\selectfont ] (n22a) at (\xo+7.0*\dx, \yo-3.0*\dy) {$\mathbf{G(\hat{\alpha}_t)\circ G_t}$\\by \eqref{eq:LieDynamics}};

\node (n21a) at (\xo+6.0*\dx, \yo-3.0*\dy) {};
\node (n22) at (\xo+5.075*\dx, \yo-2.75*\dy) {};
\node (n23) at (\xo+7.75*\dx, \yo-2.75*\dy) {};

\draw [->] (n21) -- (n22a);

\draw [->,rounded corners] (n10.south) |- ++(0.0,-0.75) |- (n22.west);

\node[circle, draw=\myFillB,font=\fontsize{13}{0}\selectfont] (n24) at (\xo+8.5*\dx, \yo-2.75*\dy) {$\mathbf{\hat{G}_{t+1}}$};

\draw [->] (n23) -- (n24);

\node[circle, draw=\myFillB,font=\fontsize{16}{0}\selectfont] (n24a) at (\xo+9.6*\dx, \yo-2.75*\dy) {$\mathbf{\hat{\Sq}_{t+1}}$};

\draw [->] (n24) -- (n24a);

\node (n25) at (\xo+3*\dx, \yo-0.25*\dy) {\textbox{{\bf Coefficient Model}}};

\node (n26) at (\xo+6.65*\dx, \yo-0.25*\dy) {\textbox{{\bf Velocity Model}}};

\begin{pgfonlayer}{background}
        \path (n26.west |- n26.north)+(-0.3,-0.45) node (a) {};
        \path (n26.west -| n26.north)+(+1.0,-2.0) node (c) {};
        \path[fill=ForestGreen!10,fill opacity=1.0,rounded corners, draw=black!20, dashed] (a) rectangle (c);           
\end{pgfonlayer}

\begin{pgfonlayer}{background}
        \path (n25.west |- n25.north)+(-0.15,-0.45) node (a) {};
        \path (n25.west -| n25.north)+(+1.0,-2.0) node (c) {};
        \path[fill=NavyBlue!10,fill opacity=1.0,rounded corners, draw=black!20, dashed] (a) rectangle (c);           
\end{pgfonlayer}

\end{tikzpicture}
    \vspace{-.1cm}
    \caption{GEM: Group Enhanced Model.}
    \label{fig:GEMscheme}
\end{figure}


\subsection{GEM Training Algorithm}\label{sec:training}
Algorithm~\eqref{alg:GEM} summarizes the  GEM training process. The algorithm receives $\{s_t ,a_t\}_{t=1}^T$ - training samples,  ($\mathcal{L}^{\alpha}_{\phi_0}$, $\mathcal{L}^{\Sp}_{\psi_0}$) - initial models, $\lambda$ - learning rate, and $K$ - number of iterations. Each training sample comprises of the state observation, $s_t$, and agent's action, $a_t$. The state observation includes the angular and translation component, $\Sq_t$, and the velocity component, $\Sp_t$. {The angular and translation  component, $\Sq_t$ is converted to $G_t$ by \eqref{eq:convertion}, which is used in the loss \eqref{eq:LieLoss}. {This transformation depends on the groups that make up the environment. For example, 2D joint angles are converted into $SO(2)$ group space. } The supplementary materials contains an walk-through of this transformation for a particular environment.}

\begin{algorithm}[t!]
   \caption{GEM Training Algorithm} \label{alg:GEM}
\begin{algorithmic}[1]
   \STATE {\bfseries Input:} $\{s_t,a_t\}_{t=1}^T$,   $\mathcal{L}^{\alpha}_{\phi_0}$, $\mathcal{L}^{\Sp}_{\psi_0}$, $\lambda$, $K$. 
   \FOR{$k = 0$ {\bfseries to} $K$}
   \STATE $\forall t\in[1,T] :$
    \STATE $\qquad (\Sq_t, \Sp_t) = s_t$
   \STATE $\qquad G_t, \Sp_t \leftarrow (\Sq_t, \Sp_t)$\hfill\COMMENT{\eqref{eq:convertion}}
   \STATE $\qquad\boxedRed{ \hat{\alpha}_t}=\boxedBlue{\mathcal{L}^{\alpha}_{\phi_k}}(G_t, a_t, \Sp_t)$
   \STATE $\qquad\;\;\hat{\Delta \Sp}_t=\boxedGreen{\mathcal{L}^{\Sp}_{\psi_k}}(G_t, a_t, \hat{\alpha}_t, \Sp_t)$
   \STATE $L^{\alpha}(\phi) = \frac{1}{T}\sum_{t=1}^T \tilde{l}\Bigl(\exp\bigl(g(\hat{\alpha}_t)\bigr)\circ G_t, G_{t+1}\Bigr)$\hfill\COMMENT{\eqref{eq:LieLoss}}
   \STATE $L^{\Sp}(\psi) = \frac{1}{T}\sum_{t=1}^T \norm{\Delta \hat{\Sp}_t- \Delta \Sp_t}_2^2$\hfill\COMMENT{velocity loss}
   \STATE $L(\phi,\psi) = L^{\alpha}(\phi) + L^{\omega}(\psi)$\hfill\COMMENT{total loss}
   \STATE $(\psi_{k+1}, \phi_{k+1})\leftarrow (\psi_k, \phi_k) - \lambda\cdot\nabla_{\{\psi, \phi\}}L(\psi_k, \phi_k)$
   \ENDFOR
\end{algorithmic}
\end{algorithm}

\section{Experiments}\label{sec:mainresult}
{
We now evaluate GEM on a standard set of continuous control dynamical systems, namely Inverted Pendulum, Inverted Double Pendulum, Reacher, Hopper, Walker, Half Cheetah, Ant, and Humanoid. These environments incorporate a diverse set of challenging geometries and motions ~\citep{todorov2012mujoco, tassa2018deepmind}, making them a suitable testbed for GEMs. 



In all the experiments, we use feed-forward multilayer perceptrons with consistent hyperparameters for the baseline model and GEM.\footnote{We use this over more advanced architectures initially to keep confounding variables to a minimum and provide a simple ablated test bed to study the effects of the geometric prior alone.  If complex architectures were initially used, it would be harder to assess the results, and additional (unrelated to geometric priors) ablation experiments would be required.} 
To make sure the benefits carry on to more advanced architectures, we also apply ensembling to both the baseline and GEM, so as to evaluate how GEM performs against the improved ensembles. We perform hyperparameter tuning over the number of hidden units, layers, and learning rates, and then select the best performing ones for both the baseline and GEM\footnote{The experiment details are provided in Supplementary Materials. The code is available at: https://tinyurl.com/GEMMBRL}. When evaluating the networks we test on online and offline data input settings. 

The experiments address the following research questions:
\begin{center}
\begin{itemize}[noitemsep,topsep=0pt]
    \item[Q1.] Does GEM achieve better long horizon prediction?
    \item[Q2.] Does GEM help better solve downstream tasks?
\end{itemize}
\end{center}

\subsection{Long horizon prediction}\label{exp:longHorizon}
Long horizon prediction is a commonly used task to test the quality of learned dynamical models~\citep{sanchez2018graph,lutter2019deep,NEURIPS2019_26cd8eca,lnn2020,NEURIPS2019_5faf461e}. 

To compare the performance of the models in this task, we first collect data in the environment using soft actor critic \citep{haarnoja2018soft}. Then we train GEMs following Algorithm \eqref{alg:GEM}, while the baseline models are trained with the original states. 

After the training, we keep both models fixed and apply action sequences with different time horizons, $t{=}1, 5, 10, 25, 50$. These sequences are derived from the same trained soft actor critic policy used before. We calculate the average step-wise error between the true state trajectory for an action sequence and the corresponding state trajectory produced by GEM model or the baseline model. To get a single number to represent the overall performance, the step-wise error is averaged over different time horizons and normalized relative to largest error between the models. 

Figure~\ref{fig:BarPlotSummary} displays the normalized relative error in all the environments averaged over five random model instantiations. It is evident that GEM outperforms its comparative baseline in the task of long horizon prediction in all the environments both as in a single model and in a model ensemble, averaging about half the error.

\begin{figure}[t!]
    \includegraphics[scale=0.12]{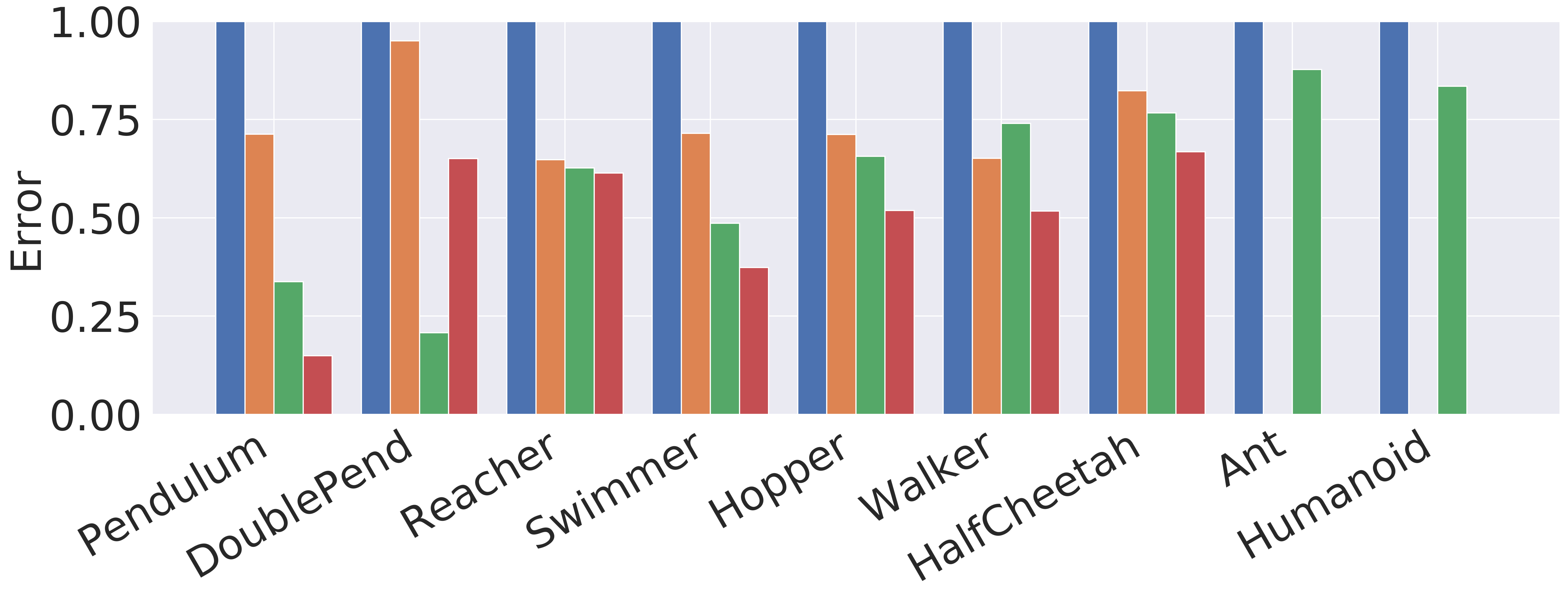}
    \caption{Relative error comparison among the baseline (blue), baseline ensemble (orange), GEM (green), and GEM ensemble (red) on continuous control tasks. The legend for this plot is the same as in Figure \ref{fig:AbsError}, given at the top right sub-plot. For Ant and Humanoid, we did not run model ensembles due to the limitation on our computational budget. Results are averaged over 5 seeds.}\label{fig:BarPlotSummary}
\end{figure}
Figure \ref{fig:AbsError} shows the detailed absolute error results for the different time horizons. We note that single step prediction performance is similar across all models. However, when scaling up the length of prediction, GEM significantly outperforms the baselines. This highlights the utility of GEM for accurate prediction of long-horizon behavior.

\textbf{Qualitative comparison:}
Figure \ref{fig:trajectories} presents samples of trajectories to demonstrate the evolution of model prediction. These images help us qualitatively understand the quantitative errors in Figure~\ref{fig:BarPlotSummary} and~\ref{fig:AbsError}. Figure \ref{fig:trajectories} shows that GEM is closer to the true model, while the baseline is much degraded, especially for the long time horizon, t=45. This observation agrees with the results presented in Figure \ref{fig:AbsError}. 

\begin{figure}[t!]
\begin{subfigure}[t]{0.48\linewidth}
    \includegraphics[scale=0.11]{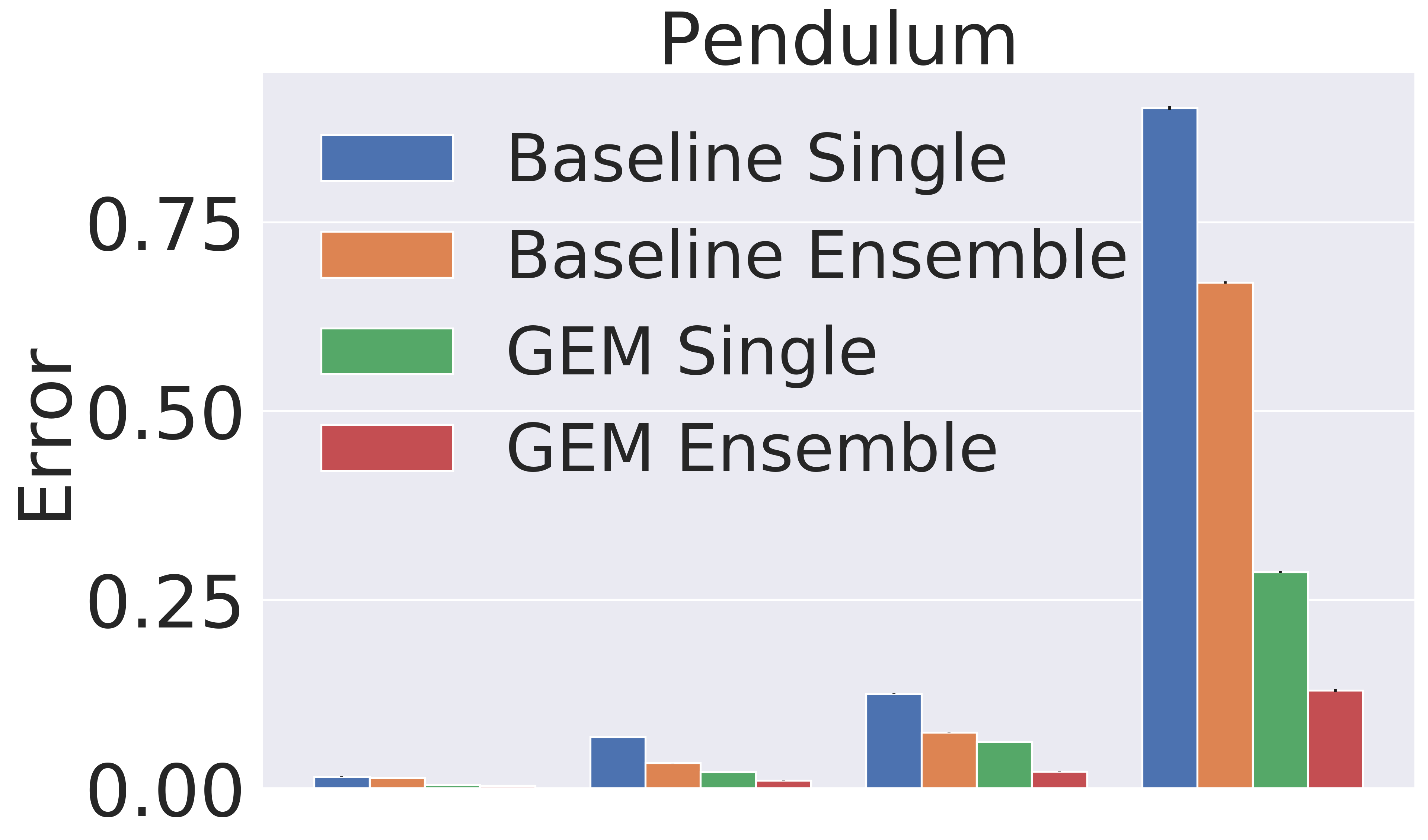}
\end{subfigure}
\;
\begin{subfigure}[t]{0.48\linewidth}
    \includegraphics[scale=0.11]{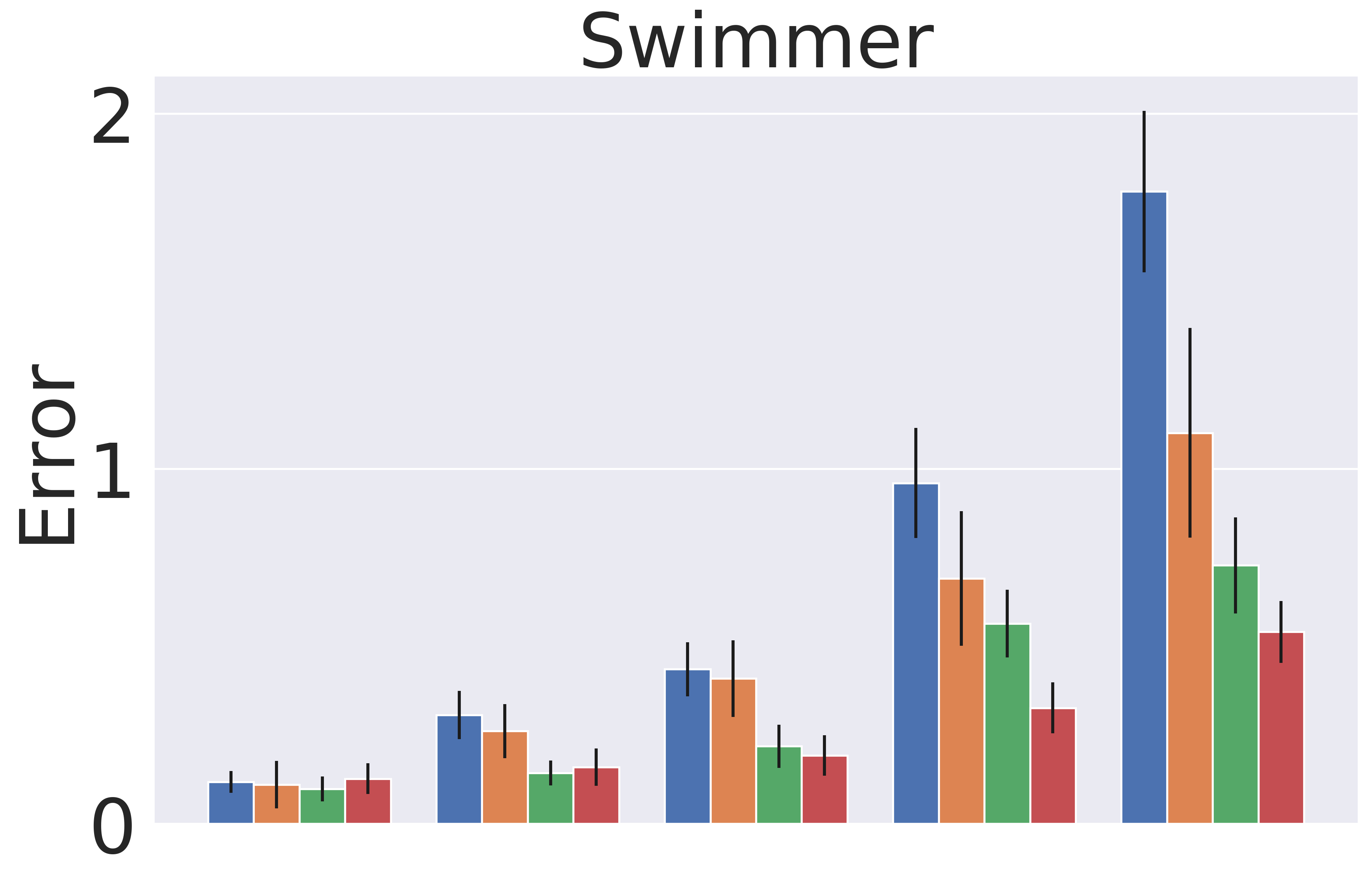}
\end{subfigure}
\\
\begin{subfigure}[t]{0.48\linewidth}
    \includegraphics[scale=0.11]{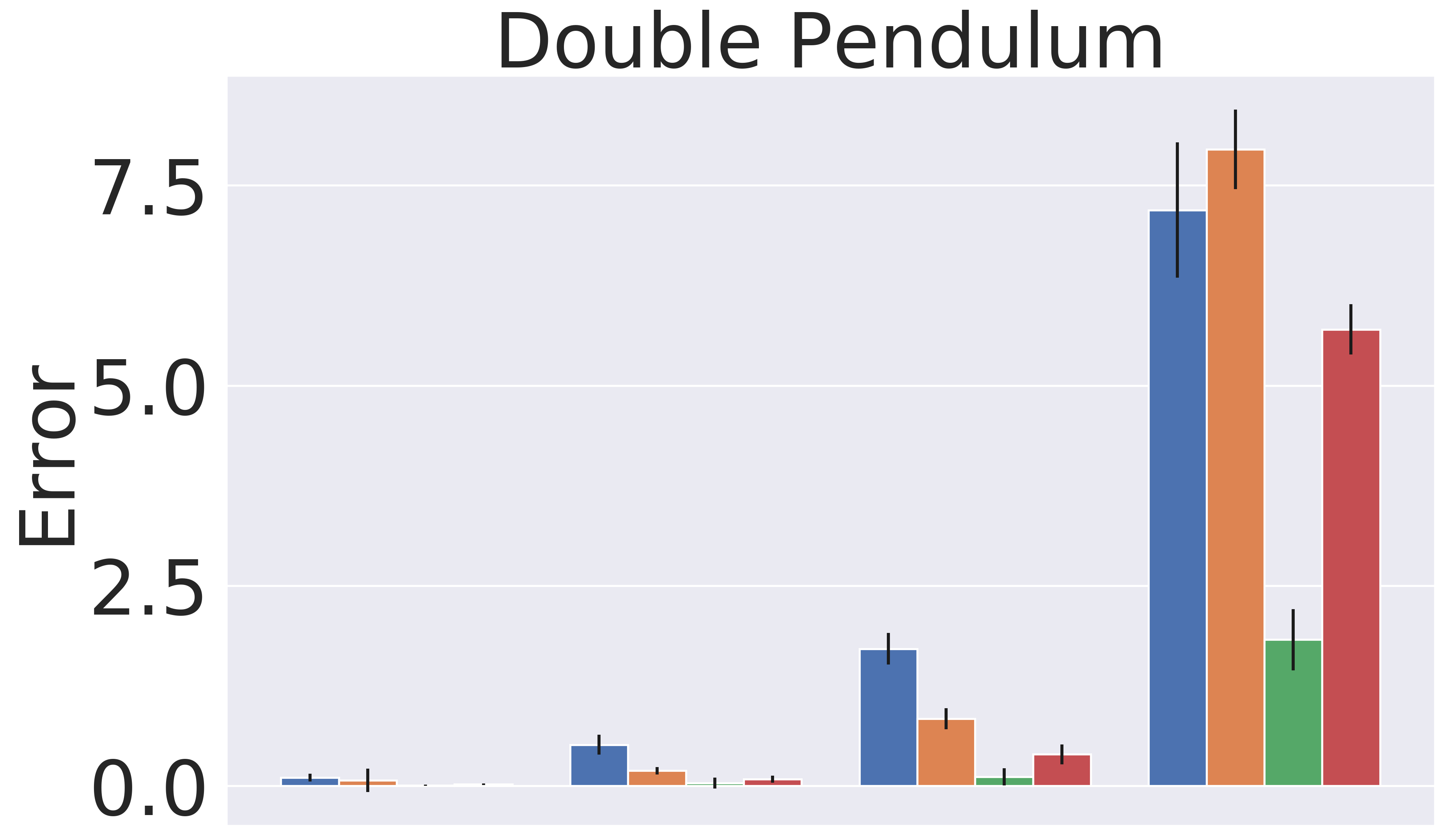}
\end{subfigure}
\begin{subfigure}[t]{0.48\linewidth}
    \includegraphics[scale=0.11]{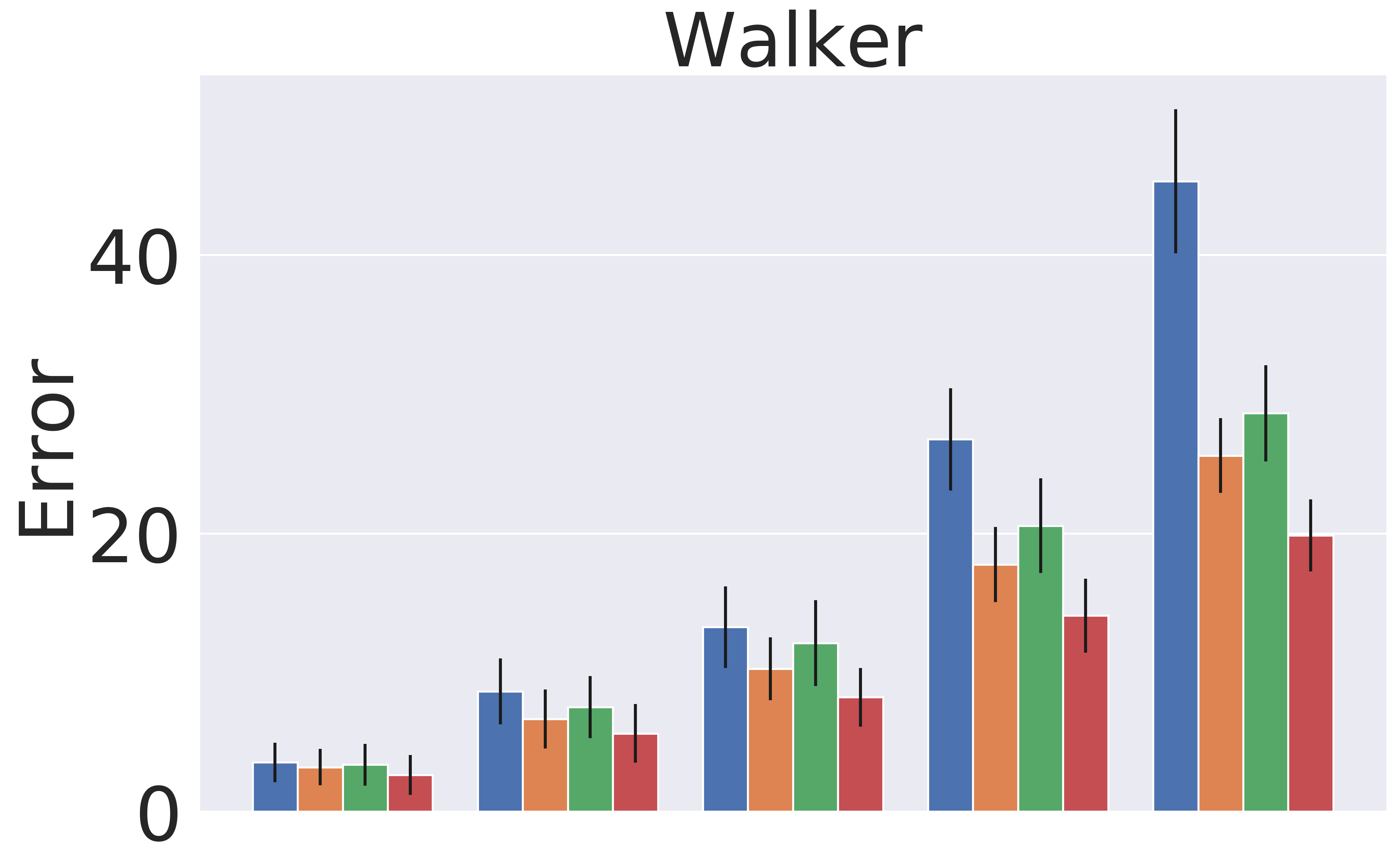}
\end{subfigure}
\\
\begin{subfigure}[t]{0.48\linewidth}
    \includegraphics[scale=0.11]{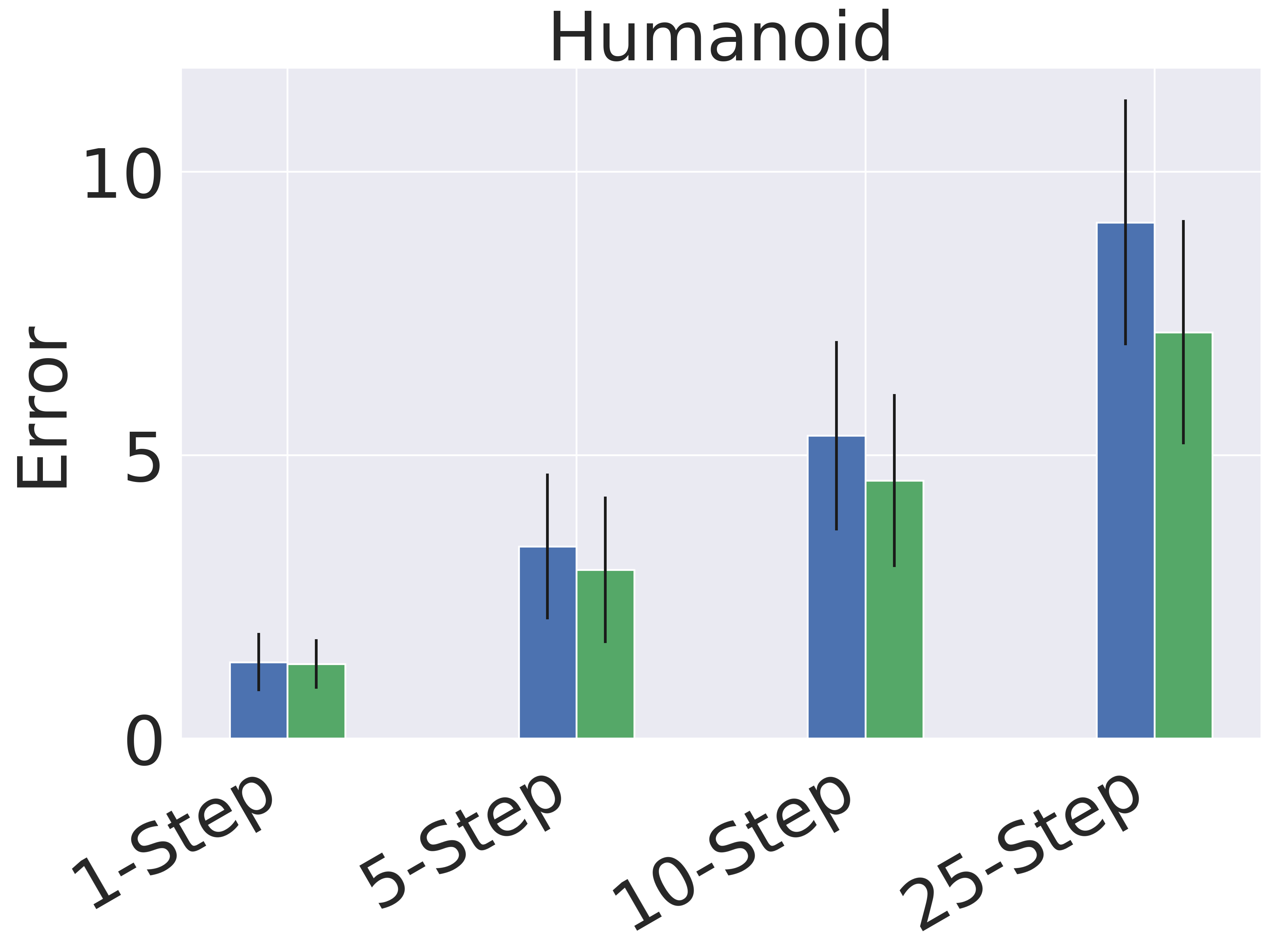}
\end{subfigure}
\;
\begin{subfigure}[t]{0.48\linewidth}
    \includegraphics[scale=0.11]{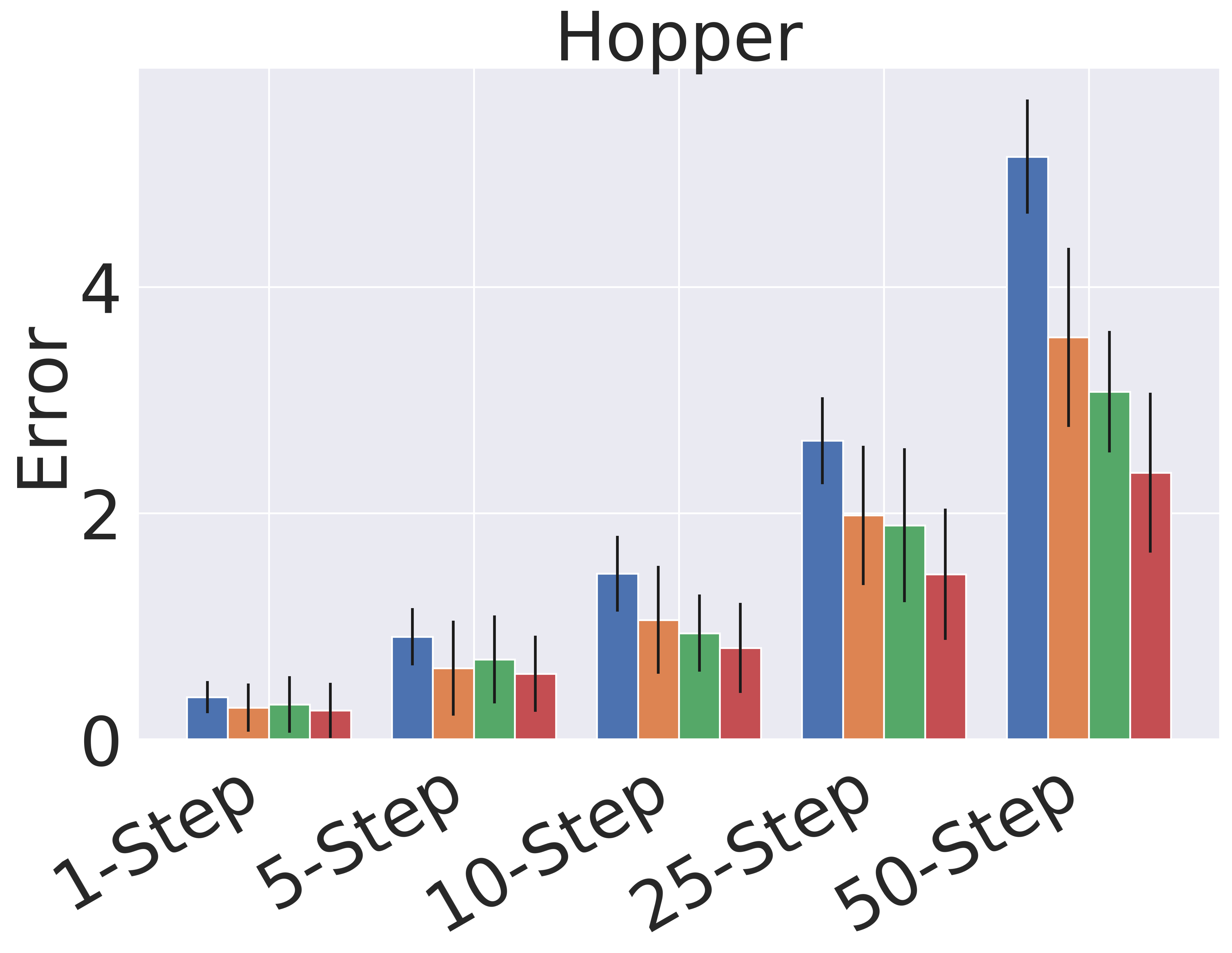}
\end{subfigure}
\caption{Absolute prediction error comparison among the baseline, baseline ensemble, GEM, and GEM ensemble on continuous control tasks at time step 1, 5, 10, and 25. Results are averaged over 5 seeds.}\label{fig:AbsError}
\end{figure}
\begin{figure}[t!]
 \begin{subfigure}[t]{0.48\linewidth}
     \includegraphics[scale=0.11]{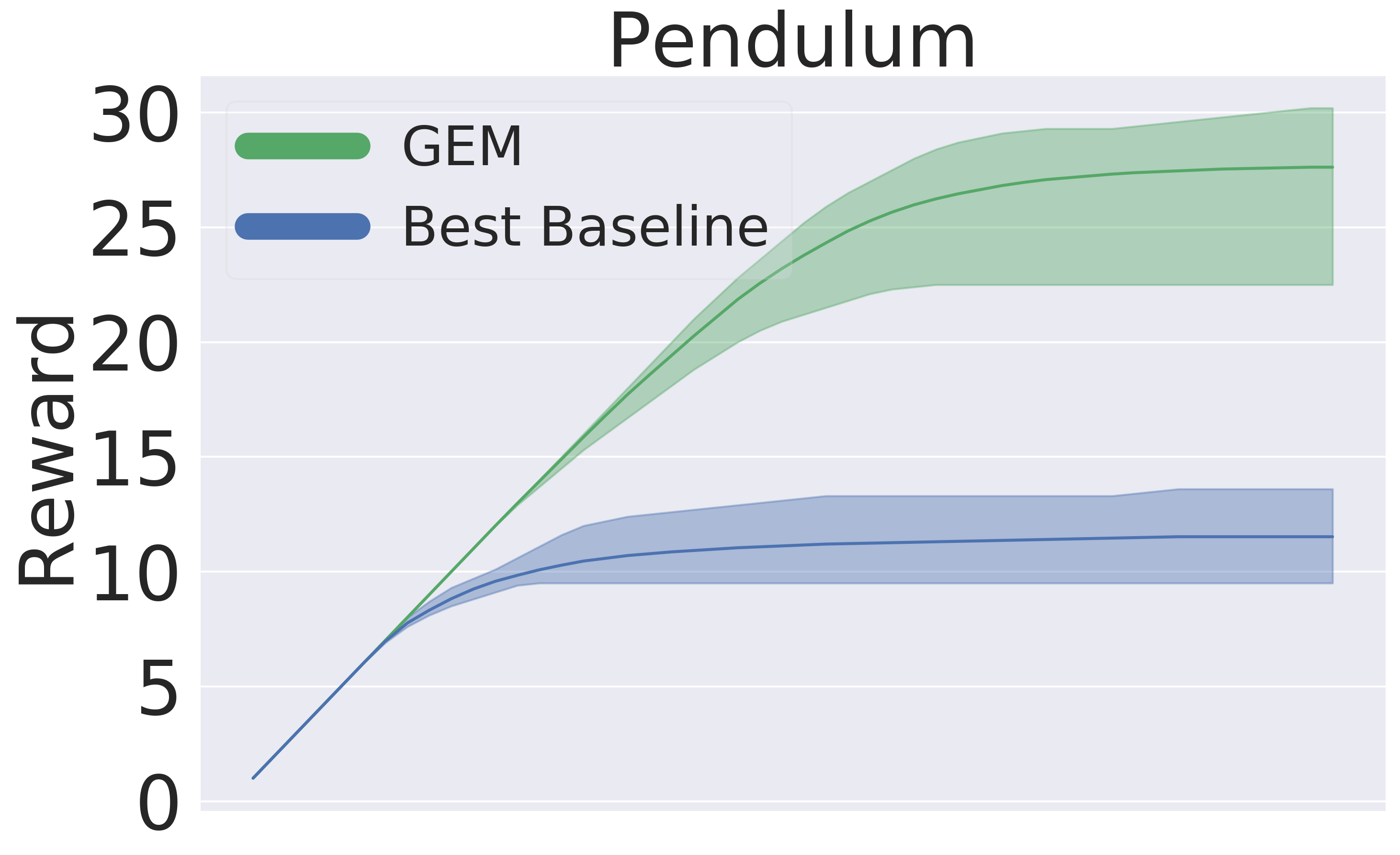}
 \end{subfigure}
 \begin{subfigure}[t]{0.51\linewidth}
     \includegraphics[scale=0.11]{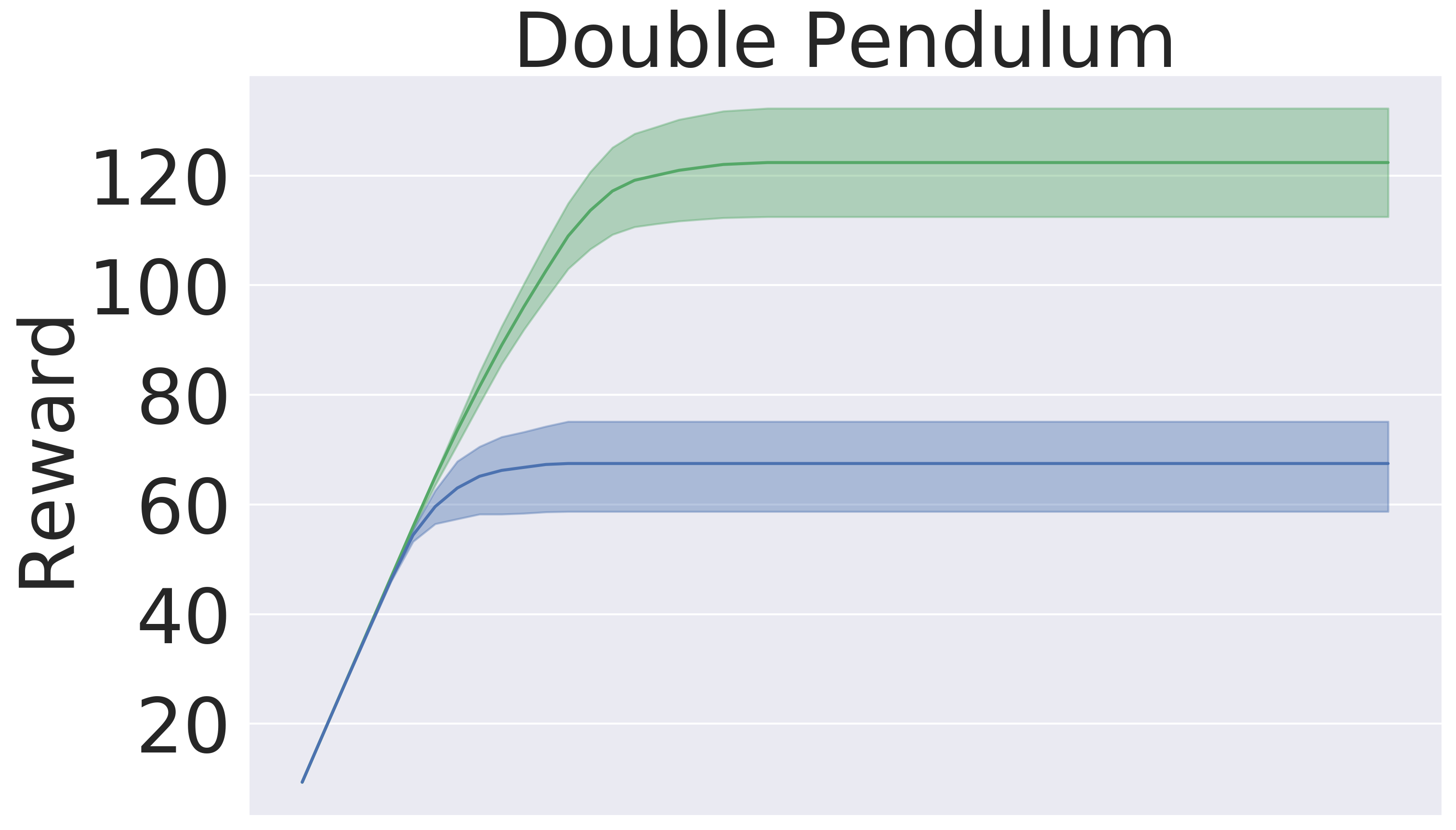}
 \end{subfigure}
 \begin{subfigure}[t]{0.49\linewidth}
     \includegraphics[scale=0.11]{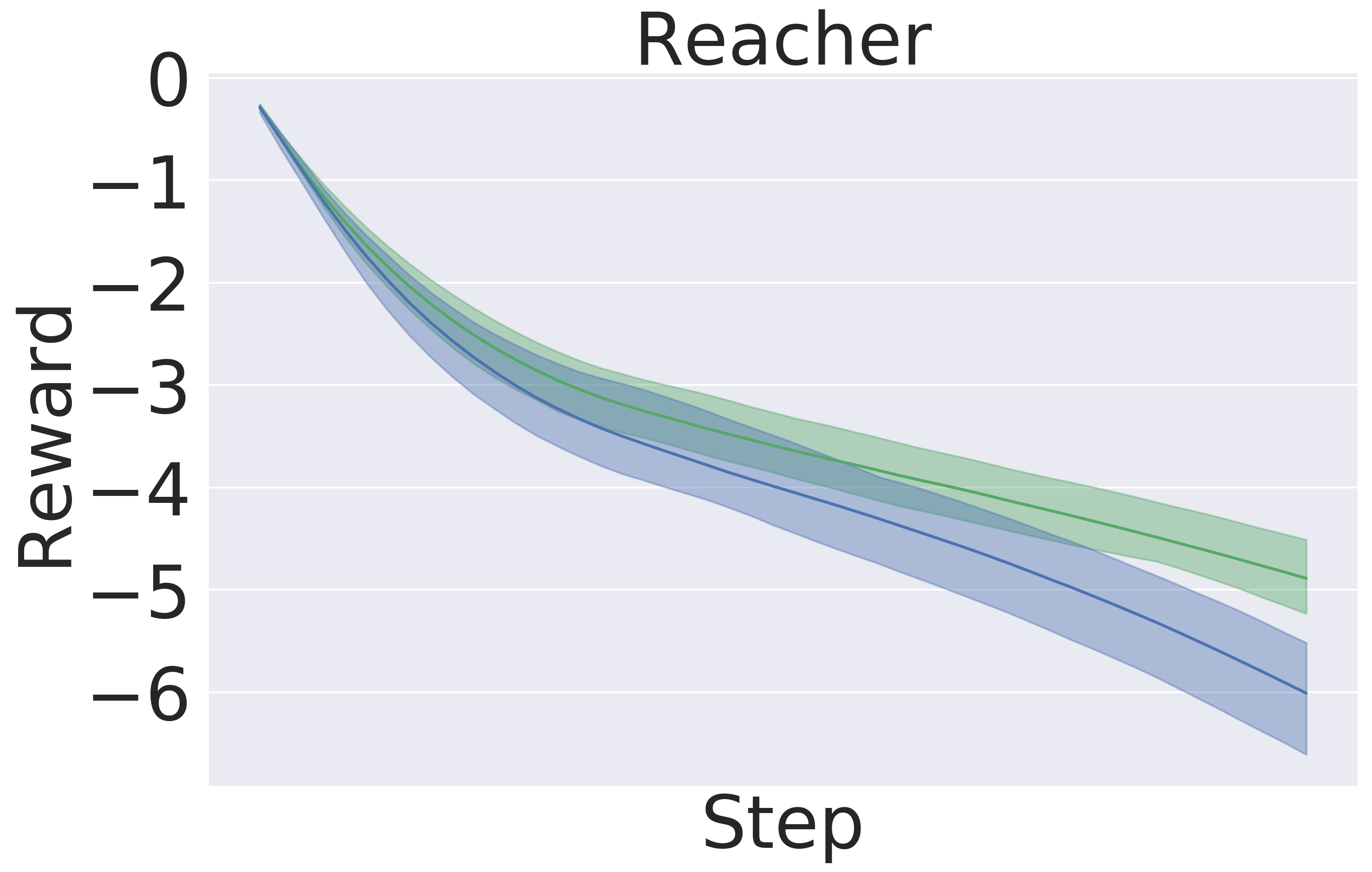}
 \end{subfigure}
\begin{subfigure}[t]{0.49\linewidth}
     \includegraphics[scale=0.11]{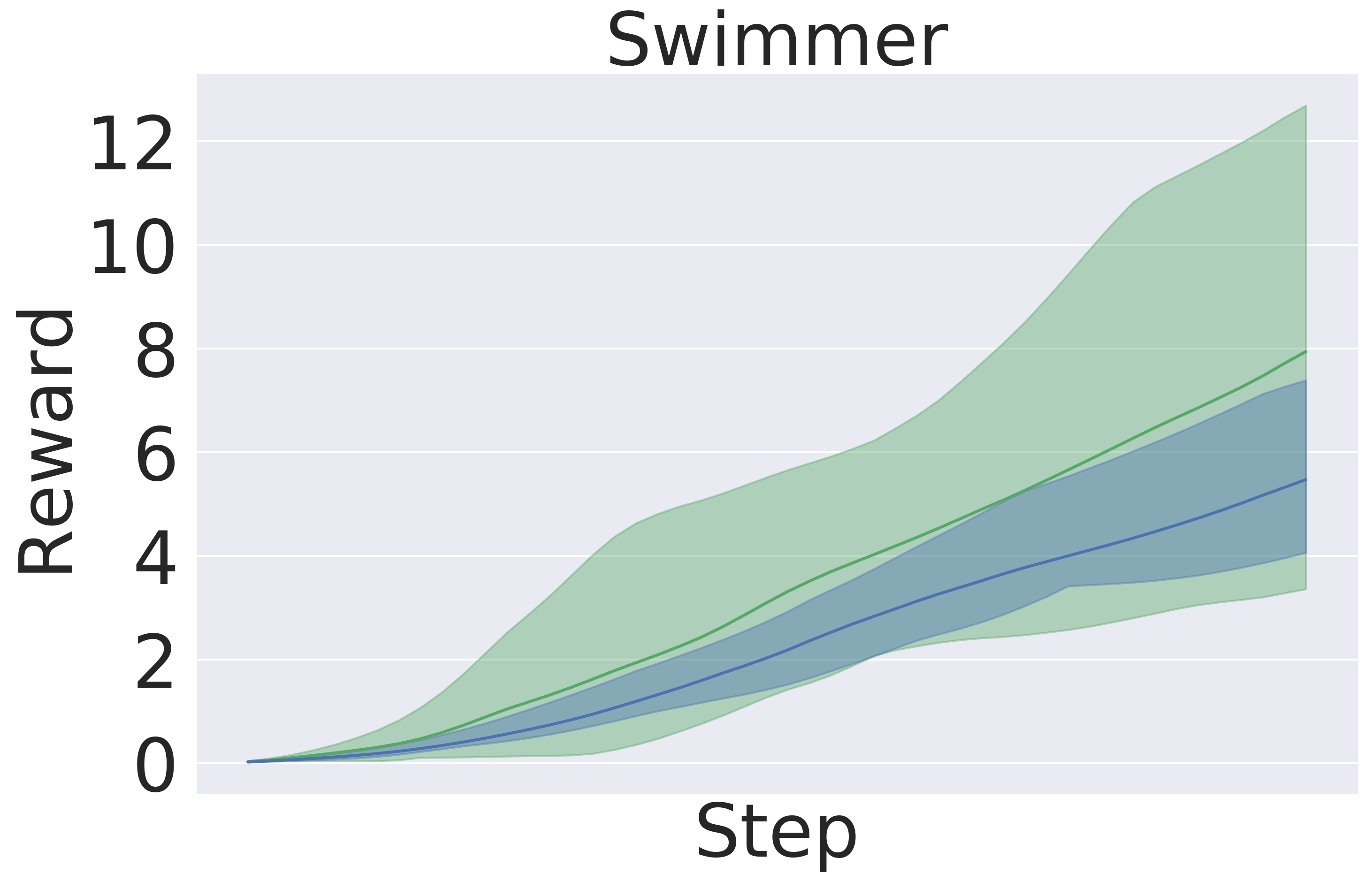}
 \end{subfigure}
 \begin{subfigure}[t]{0.49\linewidth}
     \includegraphics[scale=0.11]{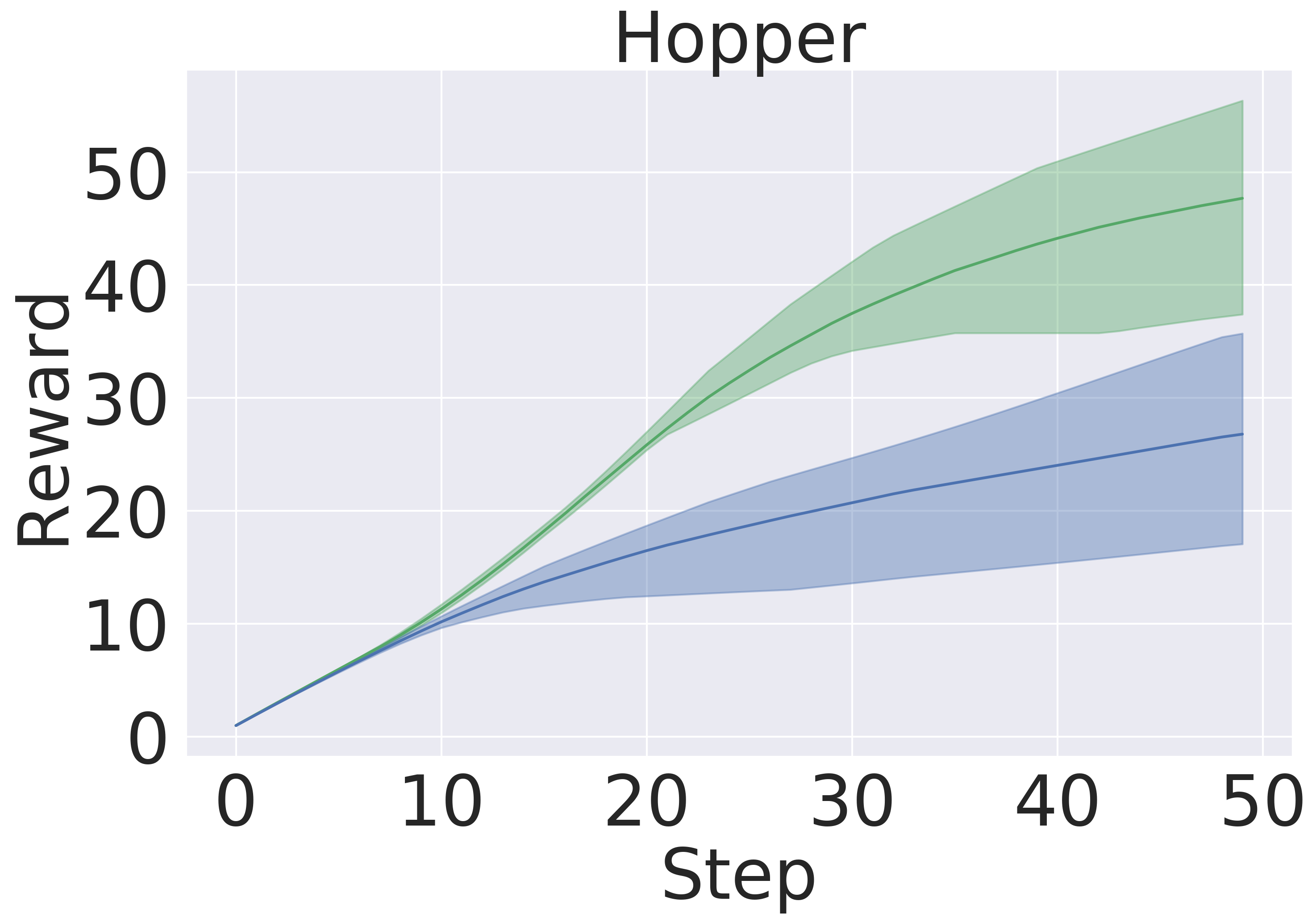}
 \end{subfigure}
 \begin{subfigure}[t]{0.49\linewidth}
     \includegraphics[scale=0.11]{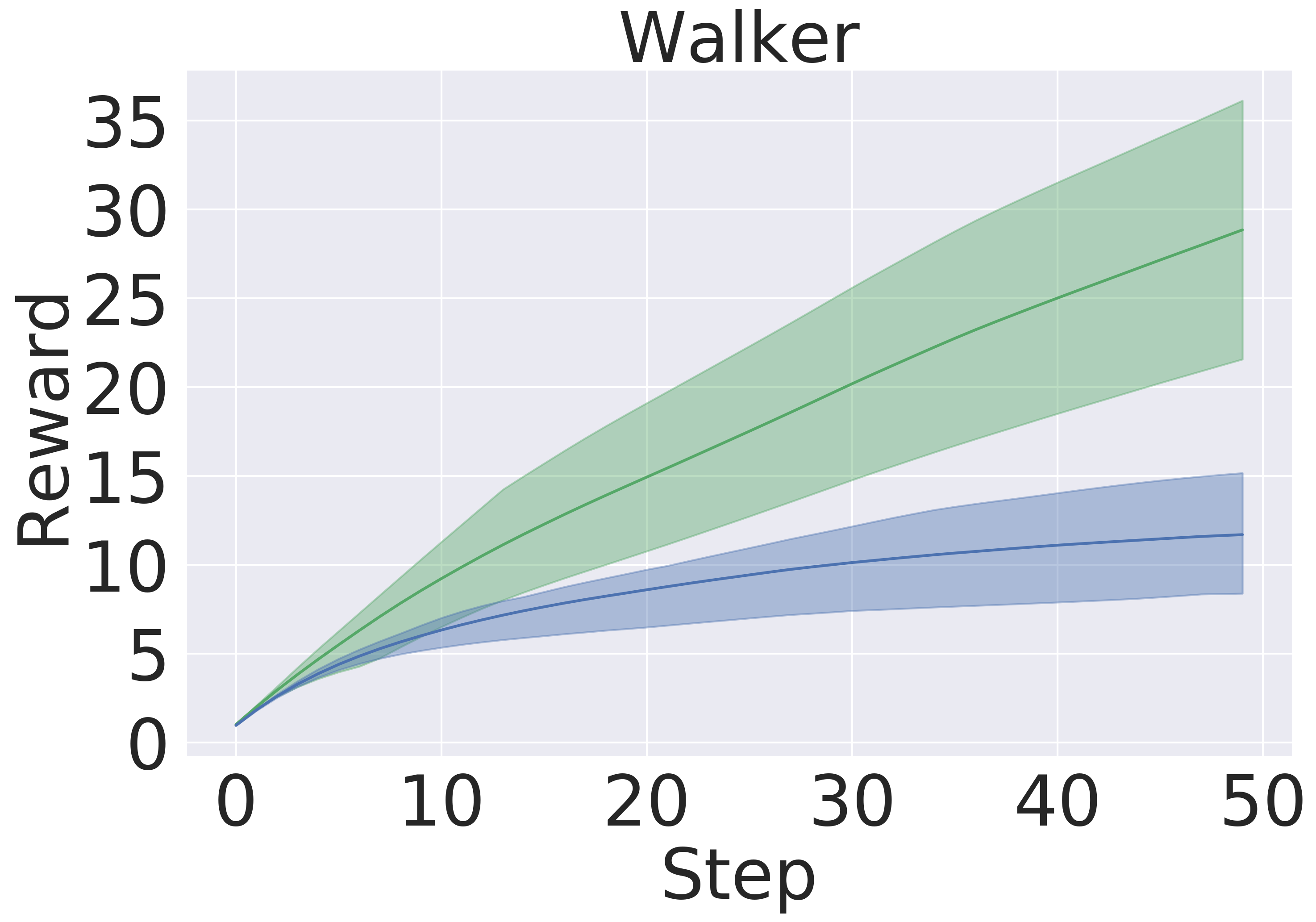}
\end{subfigure}
    \caption{Planning results from the baseline model (blue) and GEM (green). The vertical axis is the average reward and horizontal the environment step. Results are averaged over 5 random seeds with standard deviation.}\label{fig:Planning Cumulative Reward Averaged}
\end{figure}
\begin{figure}[b!]
\centering
    \includegraphics[scale=0.3]{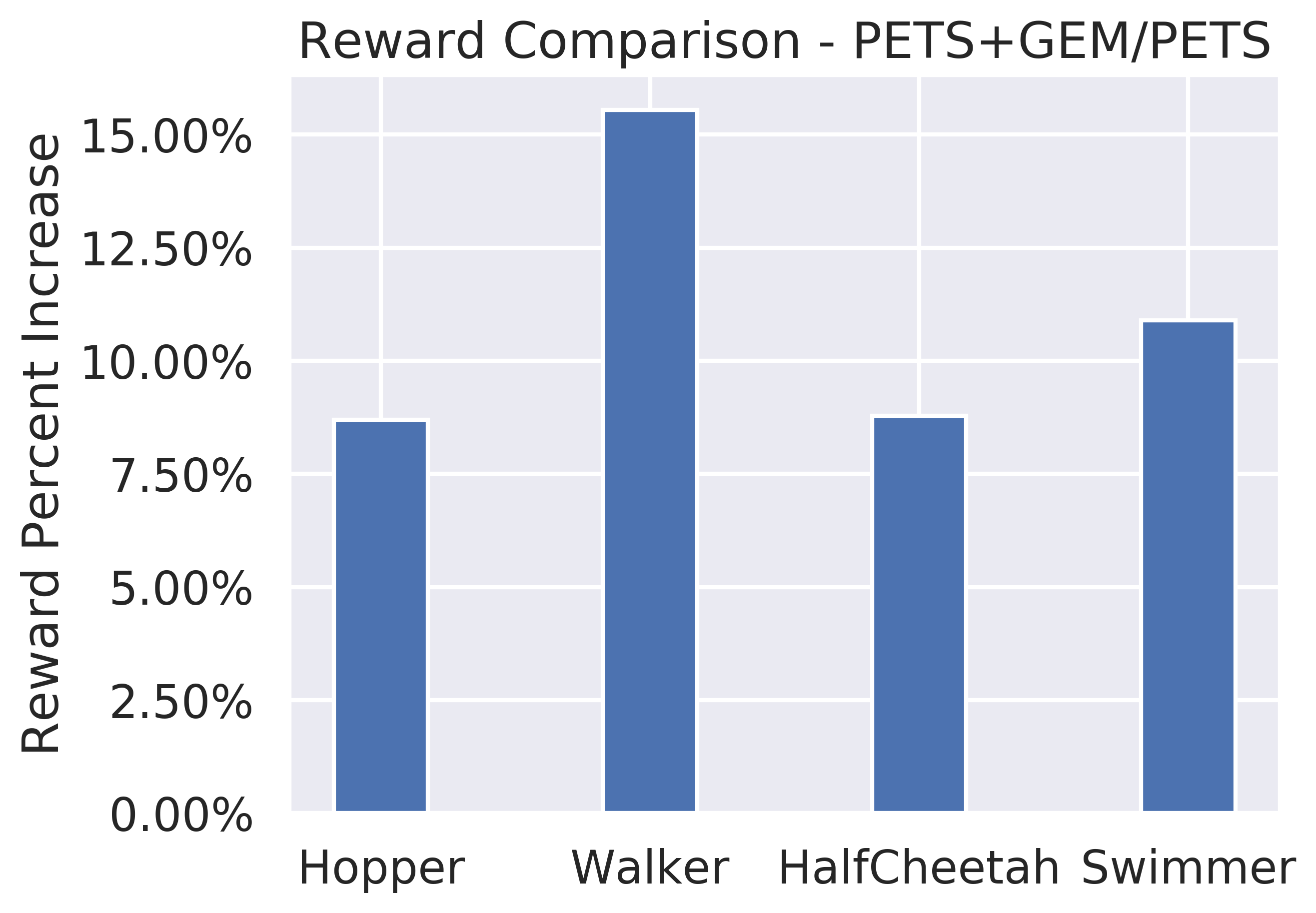}
    \caption{Improvement of PETs by augmenting it with GEM.}\label{fig:GEMPETS}
\end{figure}
\begin{figure*}[h!]
\includegraphics[scale=.39]{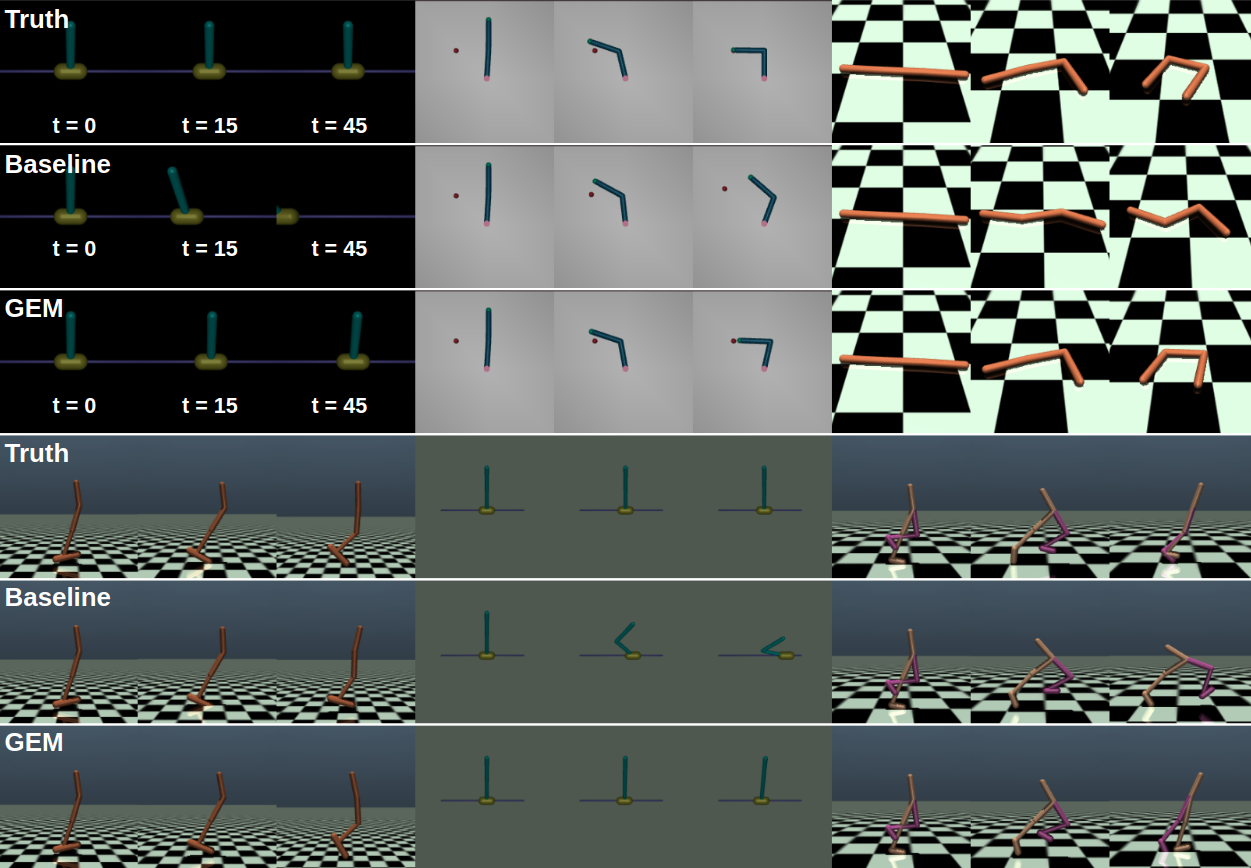}
\caption{Visualization of trajectories by the true model, the baseline model, and GEM. For each 3 by 3 block: top row from true model, middle from baseline and bottom from GEM; 1st column at $t=0$, 2nd $t=15$ and 3rd $t=45$. For example, the 3 by 3 block of images from the top left corner shows the evolution of inverted pendulum by the true model, the baseline model, and GEM, respectively, for time horizon of $t=0, t=15, t=45$ steps. We clearly observe that the baseline often fails long horizon prediction, while GEM stays close to the ground truth target.}\label{fig:trajectories}
\end{figure*}




\subsection{Planning}\label{sec:applications}


In this section we demonstrate planning experiments in order to confirm that GEM is advantageous not only for long term prediction, but also in downstream control tasks. Importantly, to decouple the effects of exploration, here, we evaluate the pre-trained models without continual updates from the environment. This property can be useful, for instance, in distributed model-based reinforcement learning, where the dynamical model is updated on a remote server, while an agent acts according to the previous model until a better model is available. 

Here, we pre-train the models in the same fashion as in Section \ref{exp:longHorizon}. Then, we use the models to plan in open loop over 50-step trajectories using an MPPI planner~\citep{tassa2018deepmind}. We apply the set of actions returned by the planner and track rewards over these 50 steps. The results are averaged over five seeds . Figure \ref{fig:Planning Cumulative Reward Averaged} presents the averaged reward over a trajectory with length 50 step within the variance due to five random initialization of initial models in pre-training. In this experiment, GEM achieves higher rewards across all tasks as well. 


\begin{figure*}[h!]
\centering
\begin{subfigure}[t]{0.18\linewidth}
    \includegraphics[scale=0.085]{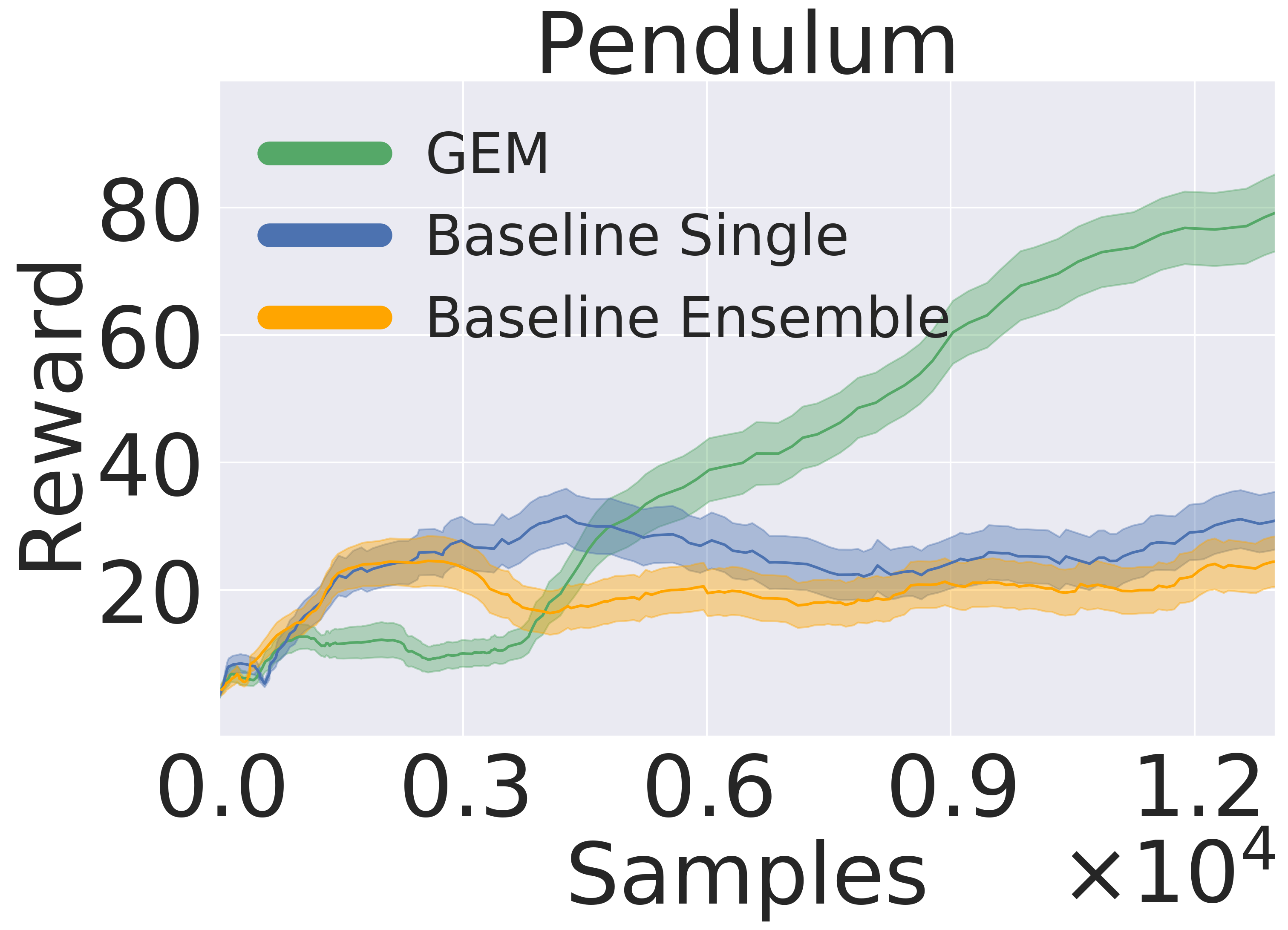}
\end{subfigure}
\;
\begin{subfigure}[t]{0.18\linewidth}
    \includegraphics[scale=0.085]{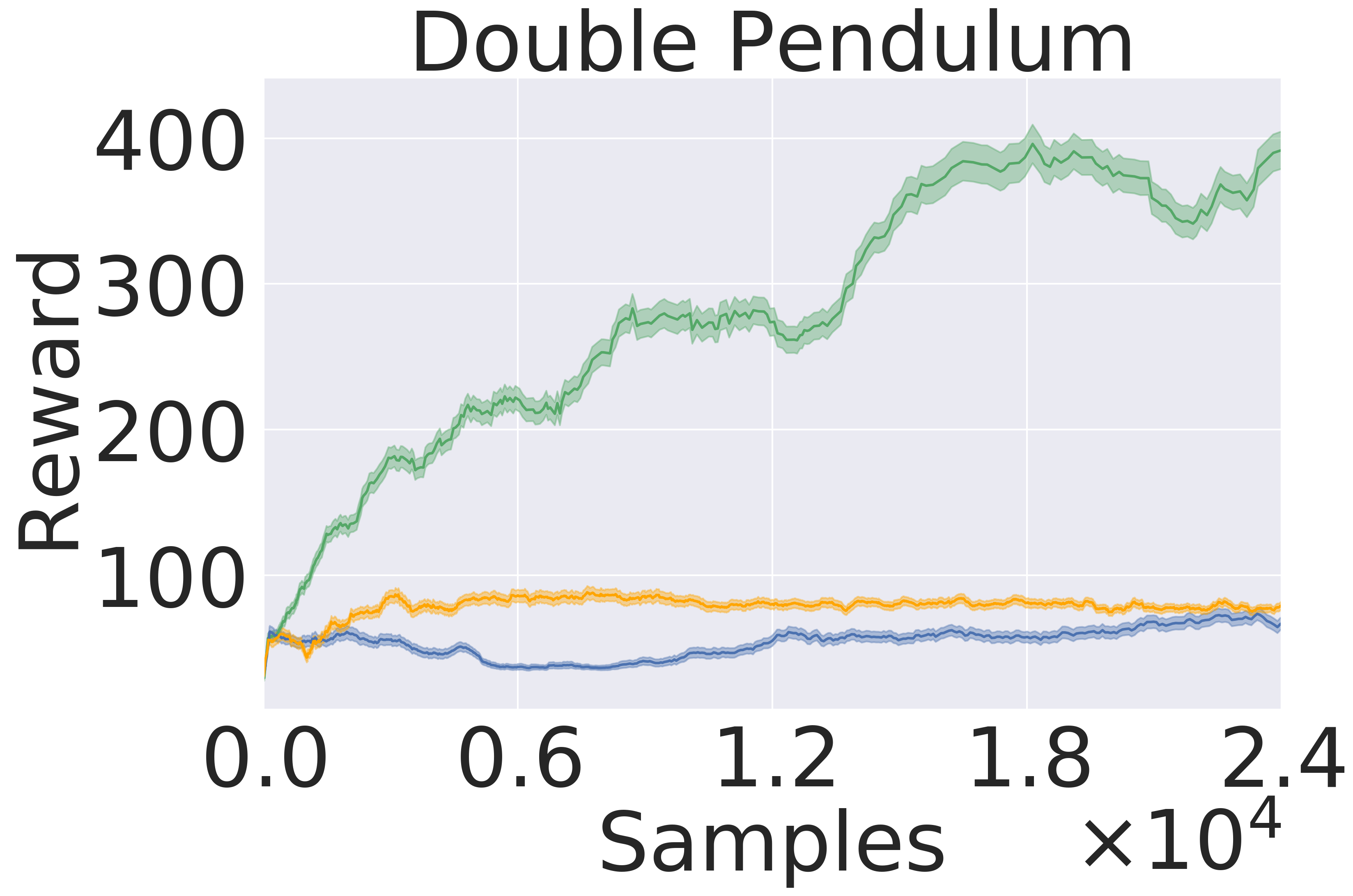}
\end{subfigure}
\;
\begin{subfigure}[t]{0.18\linewidth}
    \includegraphics[scale=0.085]{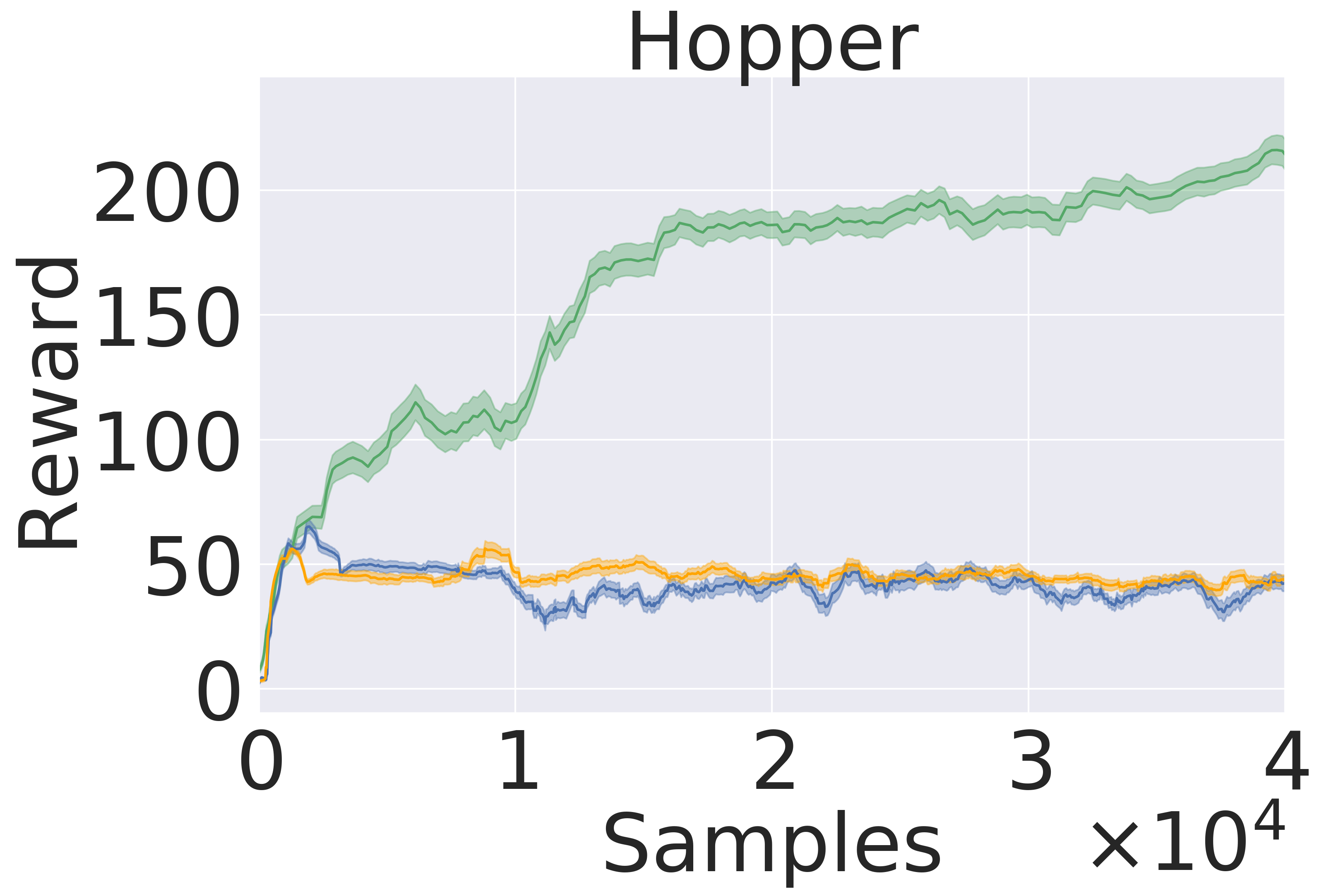}
\end{subfigure}
\;\;
\begin{subfigure}[t]{0.18\linewidth}
    \includegraphics[scale=0.085]{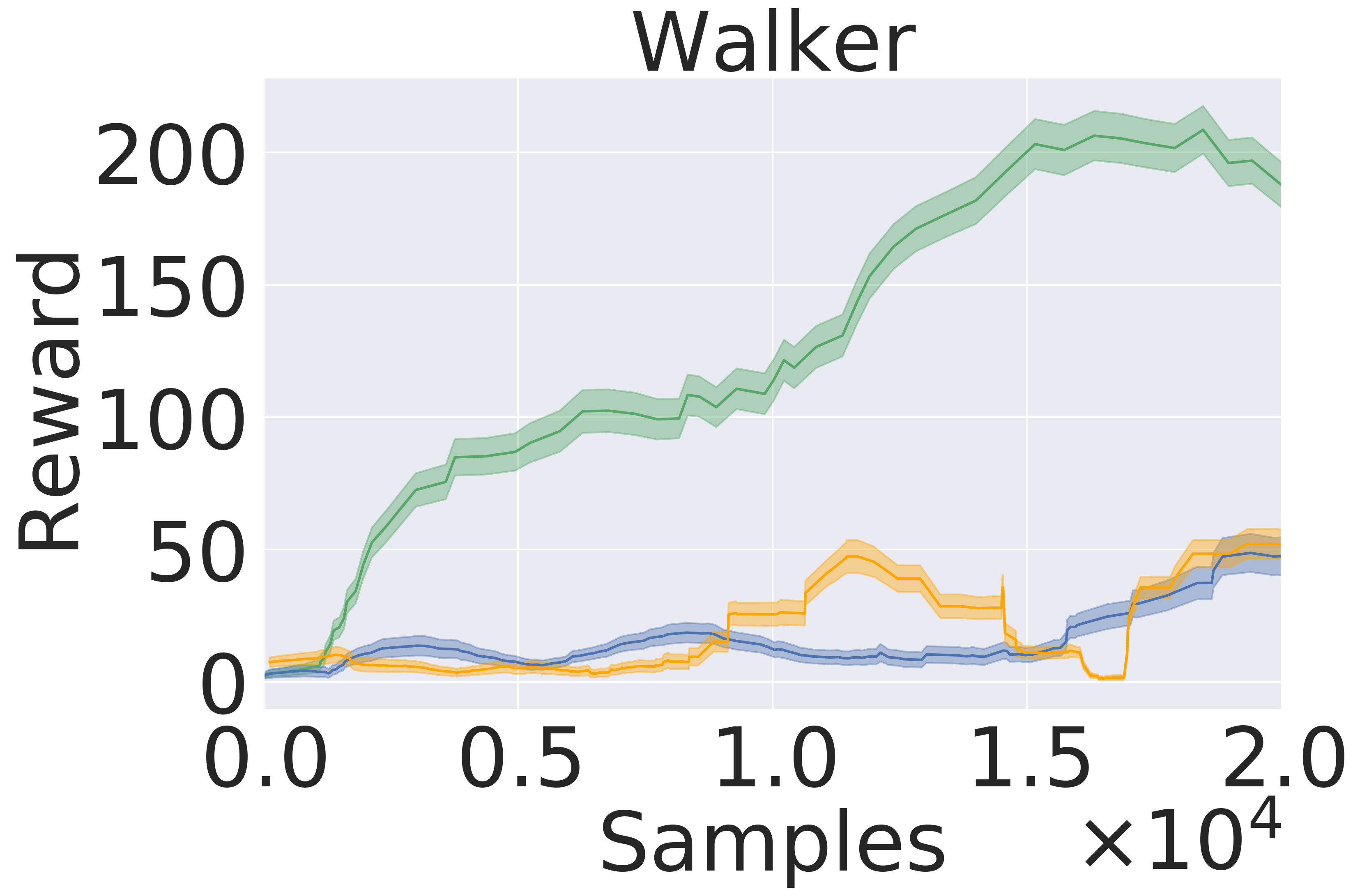}
\end{subfigure}
\;
\begin{subfigure}[t]{0.18\linewidth}
    \includegraphics[scale=0.085]{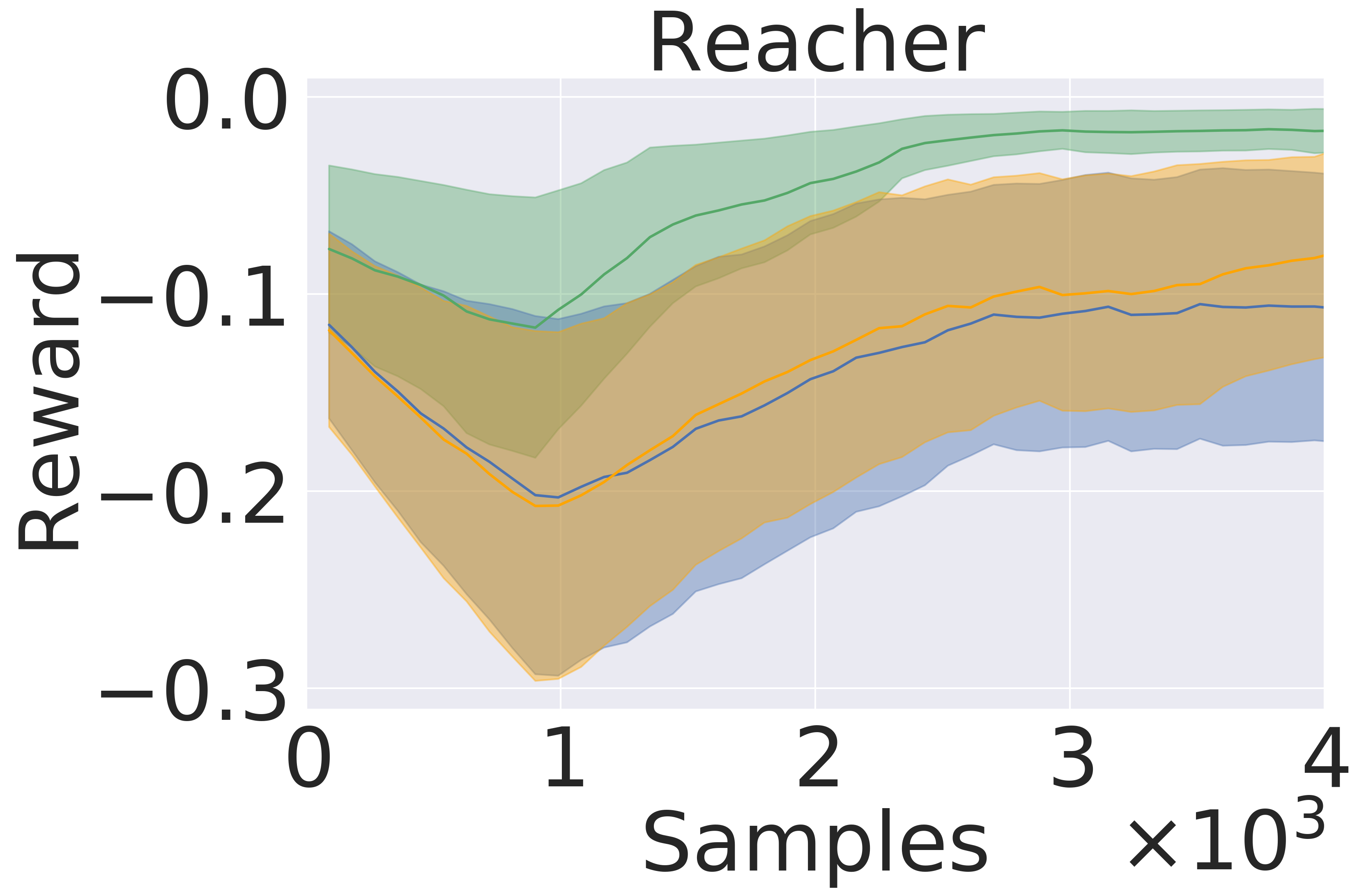}
\end{subfigure}
    \caption{We compare using the baseline (blue), baseline ensemble (orange) and GEM (green) for model-based reinforcement learning. The vertical axis is the accumulated reward and the horizontal the number of samples used. GEM significantly outperforms the baseline and the baseline ensemble, even when the latter two fail the learning entirely. The results are averaged over 5 random seeds with standard deviation.}\label{fig:MBRL Reward Averaged}
\end{figure*}

\subsection{Model-Based Reinforcement Learning}\label{sec:mbrl}
In this section we evaluate the models' performance in the setting of MBRL, where both the model and the policy are updated iteratively \citep{nagabandi2018neural, nagabandi2020deep, chua2018deep, rajeswaran2020game, Babaeizadeh2020ModelsPA}.  Importantly, observation update during execution of a trajectory happens only every fifth step in the trajectory. Figure \eqref{fig:MBRL Reward Averaged} shows that GEM achieves higher rewards in all the environment with the same amount of data. 


{
\subsection{Enhancing PETS by GEM}\label{sec:gempets}
The feed forward MLP (FFMLP) architecture used in the previous experiments is effective for testing the impact of GEM on RL and state prediction. Though it is not a SOTA architecture for model based RL. To show that GEM can improve upon SOTA  we integrate GEM to PETS \citep{chua2018deep}, which is often used as a sophisticated baseline for MBRL. This section experiments stand to confirm that the results from the FFMLP are transferable into more advanced architectures, which can be even stochastic. Figure \ref{fig:GEMPETS} shows that GEM+PETS consistently achieves higher rewards on complex environments compared to PETS alone on a similar number of trained samples and networks' parameters.}}

\section{Discussion and Conclusions}\label{sec:discussion}
In this work, we propose a framework, Group Enhanced Model, for dynamical control systems. This framework is built on the intuition that the geometric and physical properties of dynamics are naturally separable. {The former can be revealed by simple visual observation, while the later requires detailed and rigours measurement.} We use the language of Lie group theory to characterise the geometric properties of rigid body dynamics. The core in our formalism is a correspondence between 
Lie group and Lie algebra, which allows us to map points on the manifold, (e.g., rotations and translation of rigid body), to the Lie algebra vector space, and to train a dynamical model on this vector space. We experimentally assess GEM on various robotic systems by extensive comparison to the baseline models and provide answers to the research questions from Section~\ref{sec:mainresult}.

\textbf{Q1:} GEM achieves better long horizon prediction than the generic approach both as single model and as ensemble, shown in Figure~\ref{fig:BarPlotSummary} and~\ref{fig:AbsError}. In addition, Figure~\ref{fig:trajectories} qualitatively highlights that while the generic baseline can manage relatively short term prediction but fails in the long run, GEM consistently outputs close-to-truth predictions. 

\textbf{Q2:} GEM outperforms the generic baseline in downstream tasks, namely planning and model-based RL. Figure~\ref{fig:Planning Cumulative Reward Averaged} and Figure~\ref{fig:MBRL Reward Averaged} clearly exhibits the advantages of using GEM in terms of final performance. Also, GEM is more sample-efficient \textemdash reaching the same level of baseline performance significantly faster.

The current limitation of GEM is an implicit modeling of velocities through the predicted Lie algebra coefficients, rather than their direct derivation and prediction from the properties of the differential manifold. Future research could continue to explore how to tune networks with Lie algebra.



{Our advantage does not come for free, but rather we assume that we can observe \textit{what a dynamical system does} - a systems core geometry and its motions -, which, however, is often apparent, as mentioned in the abstract. We speculate that this additional knowledge will give us the advantage in comparison to learning dynamical models without knowledge of the systems geometry. }  
A rigorous research of this hypothesis is a promising new research direction. We believe GEM provides a step towards structured-RL and opens up future research on the connection between the manifold's structure and perception action cycle \cite{tishby2011information,tiomkin2017unified}.  
\bibliography{icml2021}
\bibliographystyle{icml2021}
}


{
\newpage
\newpage
\onecolumn

\pagebreak

\section*{Supplementary Materials}
In the supplementary materials, we provide a) additional details of the Lie groups used in our work, b) a working example of performing calculations with these groups, c) hyperparameters used in our experiments. The supplementary materials are not mandatory for understanding the main ideas in the paper.\footnote{The line numbers, the equation references, and the citations continue/refer those in the main paper.}

\section{Special Orthogonal Group $SO$}

A relevant example for applications in our work is the group of rotations in 3D space\textemdash\textit{Special Orthogonal group}, $SO(3)$. This section provides a detailed overview of $SO(3)$. Derived from representation theory~\cite{hall2003lie}, the elements of $SO(3)$ are orthogonal matrices with unit determinant. The corresponding Lie algebra, $\mathfrak{g}=so(3)$, is a 3-dimensional vector space. A commonly used basis for $so(3)$ is:

{\small
\begin{align}
    \!\!\!\!E_1\!\!=\!\!\begin{bmatrix}0&0&0\\0&0&-1\\0&1&0\end{bmatrix}\!\!,E_2\!\!=\!\!\begin{bmatrix}0&0&1\\0&0&0\\-1&0&0\end{bmatrix}\!\!,E_3\!\!=\!\!\begin{bmatrix}0&-1&0\\1&0&0\\0&0&0\end{bmatrix}.
\end{align}
}
From~\eqref{eq:algebraelement} in the main text, any element in $g_{\alpha}\in so(3)$ can be represented by some coefficient vector $\alpha{=}(\alpha_1, \alpha_2, \alpha_3)$, i.e. $g_{\alpha}=\alpha_1 E_1 + \alpha_2 E_2 + \alpha_3 E_e$. 
The matrix representation of $SO(3)$ allows us to conveniently use the standard matrix multiplication for the composition operator $\circ$ and the standard matrix exponential for $\exp$ in equation \eqref{eq:LieDynamics} from the main text. 

Alternatively, the elements of $SO(3)$ can be constructed from the axis-angle representation, given by a rotation axis vector, $u$, and a corresponding angle, $\theta$:
\begin{align}
    G \!= \!\cos(\theta)\!\!\cdot\!\!\mathbf{I}\!+\! \sin(\theta)\!\!\cdot\!\!(u\times u) \!+\! (1\!-\!\cos(\theta))\!\!\cdot\!\!(u\otimes u),
\end{align}
where $\mathbf{I}$, $'\times'$, and $'\otimes'$ are the identity matrix with dimension 3 by 3, the cross product, and and the outer product, respectively. Then, the angular components of a dynamical system state can be represented by rotation matrices, provided one knows the primitive geometric motions such as axes of rotations. The later can be revealed either from the hardware specification \cite{todorov2012mujoco}, or by computer vision techniques \cite{SymmetryNet,gothandaraman2020reflectional}. .

\textbf{$SO3$ and $so3$ in GEM:} given a state, we represent it by rotation $G_t\in\mathfrak{G}$\footnote{and/or translation matrices, $SE(3)$.}, and predict the Lie algebra coefficient vector, $\hat{\alpha}_{\psi}$. Thus, the predicted rotation matrix, $\hat{G}_{t+1}$, is given by Eq. \eqref{eq:LieDynamics}: $\hat{G}_{t+1}=\exp(\hat{\alpha}_{\psi})\circ G_t$. The coefficient vector, $\hat{\alpha}_{\psi}$, is represented by deep neural network with parameters $\psi$. 

{
\section{Special Euclidean Group $SE$}
For completeness, we overview the \textit{Special Euclidean group}, $SE(3)$ in Supplementary Material. These two groups characterize the primitive geometric motion of rigid body. Derived from representation theory~\cite{hall2003lie}, the elements of $SE(3)$ are essentially $SO(3)$ matrices that contain coordinates for transformations. The corresponding Lie algebra, $\mathfrak{g}=se(3)$, is a 6-dimensional vector space. A commonly used basis for $se(3)$ is:

{\small
\begin{align}
    \!\!\!\!E_1\!\!=\!\!\begin{bmatrix}0&0&0&0\\0&0&-1&0\\0&1&0&0\\0&0&0&0\end{bmatrix}\!\!,E_2\!\!=\!\!\begin{bmatrix}0&0&1&0\\0&0&0&0\\-1&0&0&0\\0&0&0&0\end{bmatrix}\!\!,E_3\!\!=\!\!\begin{bmatrix}0&-1&0&0\\1&0&0&0\\0&0&0&0\\0&0&0&0\end{bmatrix}.
\end{align}
\begin{align}
    \!\!\!\!E_4\!\!=\!\!\begin{bmatrix}0&0&0&1\\0&0&0&0\\0&0&0&0\\0&0&0&0\end{bmatrix}\!\!,E_5\!\!=\!\!\begin{bmatrix}0&0&0&0\\0&0&0&1\\0&0&0&0\\0&0&0&0\end{bmatrix}\!\!,E_6\!\!=\!\!\begin{bmatrix}0&0&0&0\\0&0&0&0\\0&0&0&1\\0&0&0&0\end{bmatrix}.
\end{align}
}
From~\eqref{eq:algebraelement} in the main text, any element in $g_{\alpha}\in se(3)$ can be represented by some coefficient vector $\alpha{=}(\alpha_1, \alpha_2, \alpha_3, \alpha_4, \alpha_5, \alpha_6)$, i.e. $g_{\alpha}=\alpha_1 E_1 + \alpha_2 E_2 + \alpha_3 E_3 + \alpha_4 E_4 + \alpha_5 E_5 + \alpha_6 E_6$. 
The matrix representation of $SE(3)$ allows us to conveniently use the standard matrix multiplication for the composition operator $\circ$ and the standard matrix exponential for $\exp$ in equation \eqref{eq:LieDynamics} from the main text. 

}

\section{An example: Reacher}
We provide an example of an environment as it goes through the GEM process.

\textbf{System description:} Reacher is a task of guiding a robotic arm to a randomized goal point. This arm is built from a limb that rotates relative to a fixed center point, the second limb rotates relative to the end of the first limb. Reacher will be built from two $SO2$ rotational groups ($SO2$ is a subset of $SO3$ discussed in \ref{sec:Preliminaries} where angles are just in two dimensions) - one group will represent the angle of the primary limb position relative to its center pivot ($\theta_1$), and the second group will represent the angle of the second limb position relative to the joint between the two limbs ($\theta_2$). The angles form the static state of reacher ($\Sq = \theta = \begin{bmatrix} \theta_1 \\ \theta_2 \end{bmatrix} $). There are also two angular velocities for each limb respectively that form the dynamic state ($\Sp = \begin{bmatrix} \Sp_1 \\ \Sp_2 \end{bmatrix}$) and two applied torques that make up the action ($\bold{a}$).

Reacher Initial Angles converted to Lie Group representation $SO(2)$, $\theta$ contains both angles and we apply each angle sequentially: \begin{align} \label{eq:angletoso2} \Sq \rightarrow \mathbf{\theta} \rightarrow \begin{bmatrix} cos(\mathbf{\theta}) & -sin(\mathbf{\theta}) \\ sin(\mathbf{\theta}) & cos(\mathbf{\theta})\end{bmatrix} = G \in SO2 \end{align}

\begin{align} \label{eq:reachercoeffmodel} \mathcal{L}_{\phi}^{\alpha}\bigg{(}\begin{bmatrix} G \\ \Sp \\ a\end{bmatrix}\bigg{)} = \hat{\alpha} = \begin{bmatrix} \hat{\alpha_1} \\ \hat{\alpha_2} \end{bmatrix} \end{align} 
The coefficients are applied to the basis matrix of the algebra for SO2, and an exponential map is taken to produce the relevant rotational matrix described in \ref{eq:expTogroup}. Then a matrix multiplication is performed for each group to predict the future position of the limb as described in equation \ref{eq:LieDynamics}.

\begin{align} \label{eq:coeffexpone} g_{\hat{\alpha}} = \begin{bmatrix} 0 & \hat{\alpha} \\ -\hat{\alpha} & 0\end{bmatrix} \xrightarrow[]{exp} \begin{bmatrix} cos(\hat{\alpha}) & -sin(\hat{\alpha}) \\ sin(\hat{\alpha}) & cos(\hat{\alpha})\end{bmatrix} = G_{\hat{\alpha}}\end{align} 

\begin{align} \label{eq:forwardstepreacher}\hat{G'} = \begin{bmatrix} cos(\theta) & -sin(\theta) \\ sin(\theta) & cos(\theta)\end{bmatrix}  \begin{bmatrix} cos(\hat{\alpha}) & -sin(\hat{\alpha}) \\ sin(\hat{\alpha}) & cos(\hat{\alpha})\end{bmatrix} \end{align}

These coefficients are first order derivatives of position, if one were to apply the coefficient $\frac{\pi}{2}$ to the first limb, it would rotate the limb 90 degrees counterclockwise. This essentially means we apply a rotational matrix of magnitude $\alpha$ radians to our group.

After predicting coefficients, the velocity model predicts future velocities. 
\begin{align} \label{eq:velocitymodelreacher} \mathcal{L}_{\psi}^{\Sp} \bigg{(}\begin{bmatrix} G \\ \Sp \\ a \\ \hat{\alpha} \end{bmatrix} \bigg{)} = \Delta \hat{\Sp'} = \begin{bmatrix}  \Delta \hat{\Sp_1'}  \\ \Delta \hat{\Sp_2'} \end{bmatrix}\end{align} 
 \\
GEMs forward prediction is complete. The loss is takes as sum of a frobenius norm difference between true delta resultant $ G'$ state and predicted delta resultant state $ \hat{G}'$ as well as a two norm difference of the true resultant $\Delta \Sp'$ and the predicted $\Delta \hat{\Sp}$. 
\begin{align} \label{eq:Loss Reacher} l_{\text{reacher}} = \| \hat{G'} -  G'\|_F^2 + \|\Delta \hat{\Sp'} - \Delta \Sp'\|_2^2\end{align}

\section{Experiment details}

The environments utilized in our experiments are from the MuJoCo control suite~\citep{todorov2012mujoco}. All experiments are run over 5 random seeds and we report the average and standard deviation. The code to run the experiments can found at https://tinyurl.com/GEMMBRL.

\subsection{Hyperparameters}
Our neural networks are multi-layer perceptions (MLPs) built using pytorch. For each task, we search over multiple hidden sizes for each model. 
For all tasks but ant and humanoid, the hidden sizes searched over are $[100], [100, 100], [100, 100, 100]$. For ant and humanoid they are $[512], [512, 512], [512, 512, 512]$. The best performing ones are used in our experiments.
For optimizer we use Adam with learning rate $5\cdot 10^{-4}$. A batch size of $10$ is used for all experiments.
Offline data samples are collected using Soft Actor Critic.
For the models trained in the offline setting, we train until convergence which take 10000 training iterations or 100000 training samples. Validation set contain 10000 samples and testing set 20000.   

{
\subsection{Ensemble Details}
For the ensemble experiments, we took 5 base models and outputted their average. They were trained in unison via total loss from all networks.

\subsection{Planning Algorithm}
We used MPPI optimizer for planning (Link to codebase: https://github.com/kzl/lifelong\_rl). Actions were taken by planning via MPC on trained dynamics model.
}

\subsection{Trajectory Prediction Experiments}
The models in this experiment are trained in an offline setting. We compare prediction power on trajectories gathered from Soft Actor Critic. In the trajectory, we pick a random starting position and take the actions from the ground truth trajectory, and move the dynamics forward based on our trained dynamics model, while keeping track of the deviation from the ground truth state. 
Then average per step error is reported for each prediction horizon in figure \ref{fig:BarPlotSummary} and \ref{fig:AbsError}. Images were sampled from these trajectories as well.

\subsection{Planning Experiments}
The models in this experiment were trained in an offline setting. Planning was done in an open loop setting: once an environments is initialized, the planner and model never receive an update from the environment besides the first initial state. Rewards are then tracked from each step in the trajectory. Both GEM and the baselines are given the reward function, which is a function of the state.

\subsection{Model Based Reinforcement Learning Experiments}
The models in this experiment were trained in an online setting. The models iterate between planning out trajectories while collecting data and training on the collected data, this was done for 300 iterations. Though, we plot samples trained on to maintain fair comparison between models. In the planning and data collection portion, the model is only updated from the environment every fifth step. Average final rewards over five random seeds are tracked relative to number of samples collected. This code is under MBRL experiments in ExperimentScripts at https://tinyurl.com/GEMMBRL.

{
\subsection{PETS Experiments}
We take the original PETS \cite{chua2018deep} code (pytorch implementation used: https://github.com/quanvuong/handful-of-trials-pytorch) and modify the input and objective of the dynamics model which is done by adding the observation to group transformation for the input, and predicting coefficients for output. The input takes in group state and velocity (similar to PETS alone). Instead of predicting direct delta state, the probabilistic ensemble is trained to predict in coefficient (algebra) space as well as delta velocity. From there, we apply \eqref{eq:LieDynamics} to get the next predicted state from the coefficients. Though there is no two step network, just one ensemble network. This was done to not change the other structures in the original PETs code. The rest of the code is the same. GEM+PET experiment appears under the PETS folder in https://tinyurl.com/GEMMBRL. Figure \ref{fig:GEMPETS} shows the comparison results for different environments.}
}


\end{document}